\documentclass[twocolumn,notitlepage,prl,superscriptaddress,longbibliography]{revtex4-2}

\usepackage{amsfonts}
\usepackage{textcomp}
\usepackage{times}
\usepackage{graphicx}
\usepackage{float}
\usepackage{latexsym,amsmath,amssymb,bm,euscript}
\usepackage{color}
\usepackage{subfigure}
\usepackage{epstopdf}
\usepackage[colorlinks=true,linkcolor=blue,citecolor=blue]{hyperref}
\usepackage{soul}
\usepackage[normalem]{ulem}
\usepackage{mathrsfs}
\usepackage{amsmath}
\usepackage{xspace}
\usepackage{natbib}
\usepackage{ulem}
\usepackage{etoolbox}
\usepackage{tikz}
\usepackage{physics}
\usepackage{chemformula}
\usepackage{natbib}
\usepackage{tikz}
\usepackage{makecell}

\newcommand{\diag}{\text{diag}}

\newcommand{\nbcp}{\ch{Na$_2$BaCo(PO$_4$)$_2$}}
\newcommand{\kcso}{\ch{K$_2$Co(SeO$_3$)$_2$}}

\graphicspath{{./Fig/}}

\begin{document}

\title{Spin Seebeck Effect of Triangular-lattice Spin Supersolid}

\author{Yuan Gao}
\affiliation{School of Physics, Beihang University, Beijing 100191, China}
\affiliation{Institute of Theoretical Physics, Chinese Academy of Sciences, Beijing 100190, China}

\author{Yixuan Huang}
\affiliation{RIKEN Center for Emergent Matter Science (CEMS), Wako, Saitama 351-0198, Japan}

\author{Sadamichi Maekawa}
\affiliation{RIKEN Center for Emergent Matter Science (CEMS), Wako, Saitama 351-0198, Japan}
\affiliation{Kavli Institute for Theoretical Sciences, University of Chinese Academy of Sciences, Beijing 100190, China}
\affiliation{Advanced Science Research Center, Japan Atomic Energy Agency, Tokai, Ibaraki 319-1195, Japan}

\author{Wei Li}
\email{w.li@itp.ac.cn}
\affiliation{Institute of Theoretical Physics, Chinese Academy of Sciences, Beijing 100190, China}
\affiliation{Peng Huanwu Collaborative Center for Research and Education, Beihang University, Beijing, China}

\date{\today}

\begin{abstract}
Using thermal tensor-network approach, we investigate the spin Seebeck effect (SSE) of the triangular-lattice quantum antiferromagnet hosting spin supersolid phase. We focus on the low-temperature scaling behaviors of the normalized spin current across the interface.  {For the 1D Heisenberg chain, we find a  negative spinon spin current in the bulk with algebraic temperature scaling; at low fields, boundary effects induce a second sign reversal at lower temperatures. These benchmark results are consistent with field-theoretical analysis.} On the triangular lattice, spin frustration dramatically enhances the low-temperature SSE, with distinct spin-current signatures --- particularly the sign reversal and characteristic temperature dependence --- distinguishing different spin states. Remarkably, we discover a persistent, negative spin current in the spin supersolid phase, which saturates to a non-zero value in the low-temperature limit and can be ascribed to the Goldstone-mode-mediated spin supercurrents. Moreover, a universal scaling $T^{d/z}$ is found at the U(1)-symmetric polarization quantum critical points. These distinct quantum spin transport traits provide sensitive  {spin current} probes for spin supersolid states in  {quantum magnets} such as Na$_2$BaCo(PO$_4$)$_2$. Furthermore, our results also establish spin supersolids as a tunable quantum platform for spin caloritronics in the ultralow-temperature regime. 
\end{abstract}
\maketitle

{\textit{Introduction.---}} 
 {Quantum magnets are fascinating correlated materials that host a diverse variety of exotic spin states and emergent phenomena.} In one-dimensional (1D) systems, spin Tomonaga-Luttinger liquid (TLL) emerges with spinon excitations~\cite{Giamarchi2003, Schlappa2012, Mourigal2013}, while higher-dimensional frustrated lattices possess even richer phases like quantum spin liquids~\cite{Anderson1973, Kitaev2006, Balents2010, Zhou2017, Broholm2020} and spin supersolids~\cite{Yamamoto2014, Sellmann2015, Wang2023SS, Mila2024Commentary, Gao2022Super, Xiang2024Nature}, etc. Recently, triangular-lattice quantum antiferromagnets Na$_2$BaCo(PO$_4$)$_2$~(NBCP)~\cite{Zhong2019,LiN2020, Lee2021, Wellm2021, Huang2022thermal, Wu2022PNAS, Chi2024,zhang2024_M, Hussain2025, Popescu2025, Sheng2025, xu2025NMR, woodland2025, Gao2022Super, Xiang2024Nature, Gao2024Dynam, Jia2024Supersolid} and K$_2$Co(SeO$_3$)$_2$~\cite{RCSO2020, chen2024phase, zhu2024continuum, xu2025KCSO, ulaga2025KCSO, Zhu2025Supersolid} have been proposed to realize spin supersolid states, which open new avenues for extreme magnetic cooling~\cite{Xiang2024Nature}. Neutron scattering studies~\cite{zhu2024continuum, chen2024phase, Gao2024Dynam, Sheng2025} and dynamical simulations~\cite{Gao2024Dynam, ulaga2025KCSO, Chi2024} have uncovered  {intriguing} magnetic excitations  {in these systems} --- from Goldstone modes and roton-like dispersions to excitation continua. 

\begin{figure}[tpb!]
\includegraphics[width=0.9\linewidth]{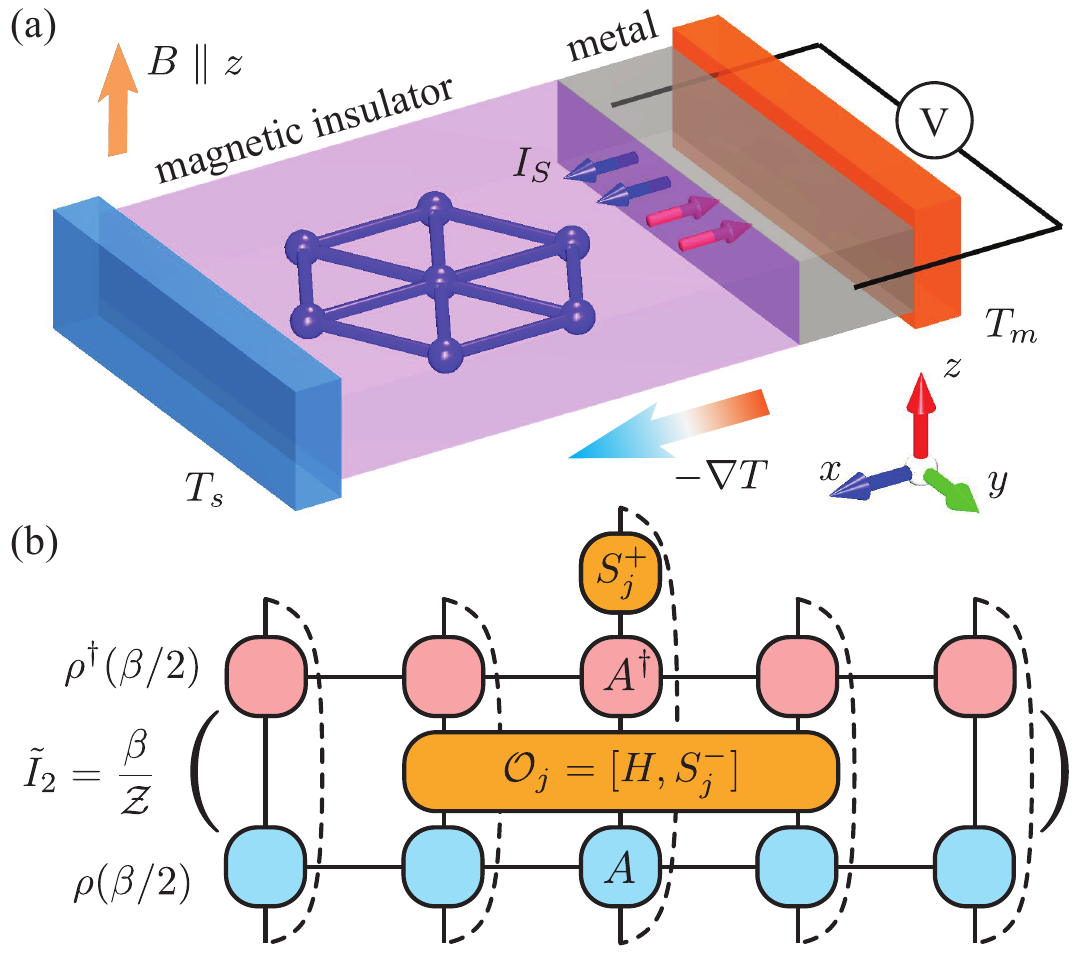}
\caption{(a)  {Longitudinal} SSE setup: the quantum magnet (triangular lattice, with temperature $T_s$) and metal substrate ($T_m$) maintain a temperature difference $\delta T = T_s - T_m$. The resulting spin current $I_S$ flows across the magnet-metal interface along the $x$-axis, parallel to the thermal gradient $-\nabla T$. Red(blue) arrow represents the positive(negative) current. A perpendicular magnetic field $B$ is applied along the $z$-axis, and the spin current is measured by the voltage $V$ along the $y$-axis through the inverse spin Hall effect in the metal substrate. 
(b) The spin current $\tilde{I}_2$ is efficiently computed by contracting the density matrix operator $\rho(\beta/2)$ with its Hermitian conjugate.  {$A$ and $A^\dagger$ are rank-4 tensors and $\mathcal{O}_j$ and $S_j^+$ are the inserted operators}.
}
\label{Fig1}
\end{figure}

An intriguing question thus emerges: Do these spin excitations lead to novel transport phenomena? In particular, the hallmark quantum transport signature --- dissipationless spin superflow --- has yet to be demonstrated in spin supersolids. Thermal conductivity measurements have been conducted on NBCP, which have produced contradictory reports of residual conductivity~\cite{LiN2020, Huang2022thermal}, highlighting the challenges in disentangling magnetic and lattice contributions~\cite{Ni2019_QSLTC, Xu2022_PRX}. On the other hand, the spin Seebeck effect (SSE), as a spin-selective transport probe~\cite{Uchida2008Nature, Uchida2010Insulator, Maekawa_SC, Adachi2013}, can offer direct access to spin current yet remains underexplored in  {frustrated} quantum magnets. The SSE is a spin analog of the Seebeck effect in magnetic compounds~\cite{ashcroft1976solid, Maekawa2004}, which reflects spin excitations by generating spin currents from thermal gradients~\cite{Takahashi2010, Xiao2010PRB, Adachi2011, Ohe2011PRB, Hirobe2017, Han2020NM}. Recently, there are theoretical studies on the sign of spin currents in spin chains~\cite{Hirobe2017, wang2025} and Kitaev magnets~\cite{Kato2025PRX}, based on spin dynamics in the ground state. However, fundamental gaps remain in understanding their temperature dependence --- especially the scaling behaviors in strongly correlated regimes. This arises from the inherent complexity in simulating SSE at finite temperature, where quantum and thermal fluctuations exhibit intriguing interplay.

In this work, we develop an efficient thermal tensor-network approach for computing the normalized spin currents and their temperature scaling, within an imaginary-time framework. We benchmark the approach on 1D Heisenberg chain  {with spinon spin current, and then apply it} to the triangular-lattice spin-supersolid system. We demonstrate that spin currents serve as effective probes for identifying distinct quantum spin states and mapping the phase diagram via their temperature dependence. In particular,  {we discover that in the spin supersolid phase, the spin current saturates} to a constant in the zero-temperature limit. Momentum-resolved analysis further demonstrates that they are mediated by dissipationless Goldstone modes --- a signature of spin supercurrent~\cite{Takei2014, Qaiumzadeh2017, Yuan2018}. Across 1D and 2D spin systems, we uncover a universal scaling $T^{d/z}$ near the polarization QCPs. Our predicted SSE features can be experimentally investigated on spin-supersolid compounds like \nbcp~\cite{Gao2022Super, Xiang2024Nature} and \kcso~\cite{zhu2024continuum,chen2024phase}.

\begin{figure}[]
\includegraphics[width=1\linewidth]{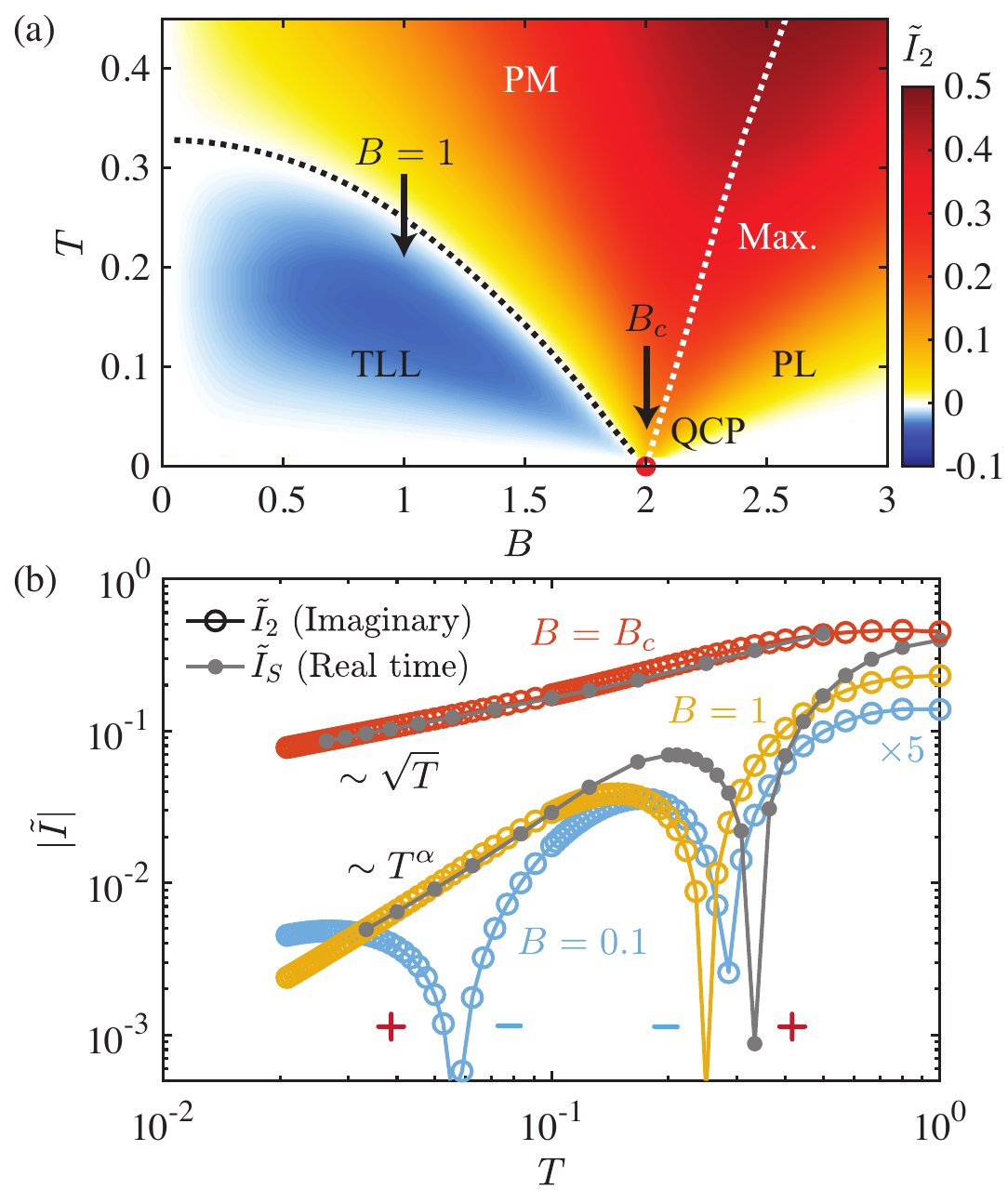}
\caption{Benchmarks on normalized spin current in 1D Heisenberg chain ($L=128$, $D=500$). 
(a) The simulated spin current $\tilde{I}_2$, where the black dotted line marks the sign reversal, and the white dotted line locates the maximum of $\tilde{I}_2$ under a fixed field. The red dot labels the QCP at $B_c=2$, separating the TLL and polarized (PL) phases. 
(b) presents the temperature dependence of spin currents calculated through real-time dynamics ($\tilde{I}_{S}$, with $t_{\rm max}=40$, $D=500$  {and $j=64$}) and imaginary-time correlations ($\tilde{I}_2$,  {$j\in[33,96]$ for $B=1, B_c$ and $j\in[1,2]$ for $B=0.1$}).  {At low temperatures, we observe algebraic spin current $\tilde{I}_{S,2} \sim T^{\alpha}$ with $\alpha \simeq 1.59(2)$ for $B=1$ (TLL phase) and $\tilde{I}_{S,2} \sim \sqrt{T}$ at $B=B_c$ (QCP).} Given the undetermined prefactors in simulated spin currents, we shift the $\tilde{I}_{S}$ data to align with the low-temperature $\tilde{I}_2$.  {The slight shift of the sign-reversal temperature is ascribed to their different kernel functions, namely, $k(\beta\omega)$ versus $k^2(\beta\omega)$. For a small magnetic field $B=0.1$, the boundary contributions from first two sites exhibit an additional sign reversal at a lower temperature.}
}
\label{Fig2}
\end{figure}

\begin{figure*}[]
\includegraphics[width=1\linewidth]{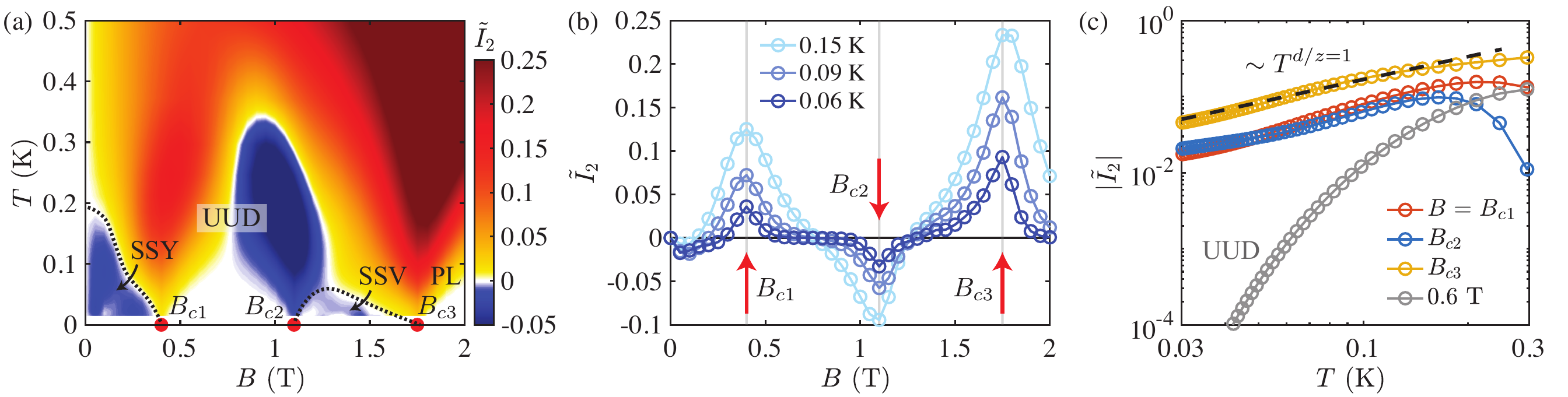}
\caption{(a) Simulated spin current $\tilde{I}_2$ of TLAF model (6$\times$18 cylinder, $D=3000$) for the compound NBCP~\cite{Gao2022Super}. Three QCPs $B_{c1,2,3}$ (red dots) separate supersolid-Y (SSY), up-up-down (UUD), supersolid-V (SSV), and the PL phases. Dashed lines indicate schematic phase boundaries of SSY and SSV determined from the sign reversal in $\tilde{I}_2$. (b) Isothermal $\tilde{I}_2$ cuts reveal three QCPs at $B_{c1} \simeq 0.4$~T, $B_{c2} \simeq 1.1$~T, and $B_{c3} \simeq 1.75$~T (vertical gray lines), with prominent peaks or dips. (c) $\tilde{I}_2 \sim T^{d/z}$ with $d/z=1$ (black dashed line) at QCPs ($B_{c1,2,3}$). The exponentially decaying $\tilde{I}_2$ within the UUD phase ($B=0.6~{\rm T}$) is also plotted as a comparison.}
\label{Fig3}
\end{figure*}

\textit{Thermal tensor-network calculations of spin current.---}
Here we consider the XXZ Heisenberg model under a magnetic field, i.e., $H=H_0- B \sum_i S_i^z$, where 
\begin{equation}
H_0 = \sum_{\langle i,j \rangle} \frac{J_{xy}}{2}(S_i^+ S_j^- + S_i^- S_j^+) + J_z S_i^z S_j^z.
\label{Eq:H}
\end{equation}
The couplings $J_{xy},J_z > 0$ represent the nearest-neighboring antiferromagnetic exchange, and $B$ is the external field. As shown in Fig.~\ref{Fig1}(a), the spin current $I_S$ across the magnet-metal interface is driven by temperature gradient and expressed as $I_S = -A\tilde{I}_S\delta T$, where $A$ denotes a material-dependent constant and $\delta T \equiv T_s - T_m$ represents the temperature difference across the interface. Derived through non-equilibrium Green's function formalism~\cite{Adachi2011, Adachi2013, Masuda2024, Supplementary}, the normalized spin current $\tilde{I}_S$ takes the form
\begin{equation}
\tilde{I}_S = \int_{-\infty}^\infty ~ d\omega ~ k^2(\beta\omega) \, \mathrm{Im}[\chi^{-+}_\mathrm{loc}(\omega)], 
\label{Eq:IS}
\end{equation}
with kernel function $k(x) = x/\sinh(x/2)$, where $x\equiv \beta\omega$ and $\beta \equiv 1/T$. The local dynamical susceptibility $\chi^{-+}_\mathrm{loc}(\omega)$ is the central quantity of interest for determining the spin current. One approach for $\tilde{I}_S$ involves computing $\mathrm{Im}[\chi^{-+}_\mathrm{loc}(\omega)]$ in the ground state~\cite{Kato2025PRX, wang2025}, while incorporating temperature influences solely through the kernel function $k^2(\beta \omega)$~\cite{Supplementary}. 

To accurately account for the temperature dependence of ${\rm Im}[\chi^{-+}_\mathrm{loc}(\omega)]$, we  {exploit the thermal tensor-networks~\cite{tanTRG2023,Li2011,ChenSETTN2017,Chen2018} (see Appendix)}, based on the imaginary-time framework,  {to achieve} efficient and accurate calculations of SSE in both 1D chains and 2D frustrated lattices. Given the analyticity of ${\rm Im}[\chi^{-+}_{\rm loc}(\omega)]$ near $\omega = 0$, we expand ${\rm Im}[\chi^{-+}_{\rm loc} (\omega)] = \sum_{n=1}^\infty f_n \frac{\omega^n}{n!}$. The even parity of the kernel function $k(\beta\omega)$ selects $f_2$ as the leading term, resulting in the dominant contribution $\tilde{I}_S \sim f_2/\beta^3$, which accurately captures the low-temperature scaling for $\beta \omega \lesssim {O}(1)$. On the other hand, we express the local imaginary-time correlation function as
$\frac{\partial}{\partial \tau} \langle S^-_j(\tau) S^+_j \rangle \Big|_{\tau = \beta/2} = \frac{1}{2 \beta \pi} \int_{-\infty}^\infty d\omega \, k(\beta\omega) \, {\rm Im} [\chi_{\rm loc}^{-+}(\omega)] = \frac{f_2}{\beta^4} +O(\frac{1}{\beta^6}) \sim \frac{\tilde{I}_S}{\beta}$,
where ${\rm Im}[\chi_{\rm loc}^{-+}(\omega)]$ is also expanded up to second order. Therefore, the normalized spin current can be calculated in the low-temperature regime  {via the imaginary-time approximation}~\cite{Supplementary}
\begin{equation}
\tilde{I}_2 = \beta \langle \mathcal{O}_j(\frac{\beta}{2}) S_j^+ \rangle_\beta  {\, \sim \tilde{I}_S}.
\label{Eq:I2}
\end{equation}
Here $\mathcal{O}_j= [H_, S_j^-]$ is a local operator satisfying $\frac{\partial}{\partial \tau} \langle S^-_j(\tau) S^+_j \rangle \Big|_{\tau = \beta/2} = \langle \mathcal{O}_j(\beta/2) S_j^+ \rangle_\beta$. In practice, using the tangent-space tensor renormalization group (tanTRG) method~\cite{tanTRG2023}, we prepare the thermal density matrix $\rho(\beta/2) = e^{-\beta H/2}$ in the matrix product operator form. Subsequently, the imaginary-time correlation function and thus $\tilde{I}_2$ can be obtained through the tensor-network contraction scheme depicted in Fig.~\ref{Fig1}(b). 

\textit{Benchmarks on 1D Heisenberg spin chain.---}
We begin by analyzing the isotropic Heisenberg spin chain ($J_{xy} = J_z = 1$), where spin currents are computed using our thermal tensor network method for a finite-size chain and  {averaged in the bulk}. In such 1D spin chains, we observe excellent agreement between $\tilde{I}_2$ and $\tilde{I}_S$, with the latter obtained via computationally intensive real-time evolution of finite-temperature quantum states~\cite{Supplementary}. For 2D systems, $\tilde{I}_2$ remains capable of accurately capturing spin currents despite the prohibitive computational cost of calculating $\tilde{I}_S$.

In Figure~\ref{Fig2}(a), we show the contour plot of $\tilde{I}_2$, which reveals a characteristic sign reversal that locates the crossover between the low-temperature TLL (negative) and the high-temperature paramagnetic regimes (positive). To validate the $\tilde{I}_2$ results, we also perform real-time calculations of $\tilde{I}_S$ through Eq.~(\ref{Eq:IS}) as a benchmark. The real-time dynamical correlation function and corresponding local susceptibility ${\rm Im}[\chi_{\text{loc}}^{-+}(\omega)]$ are evaluated using tensor network method that combines finite-temperature tanTRG~\cite{tanTRG2023} and time-dependent variational principle approach for real-time dynamics~\cite{TDVP2011,TDVP2016,Supplementary}. 
 
As shown in Fig.~\ref{Fig2}(b), we find both $\tilde{I}_S$ and $\tilde{I}_2$ exhibit consistent temperature scaling at low temperature ($T \lesssim 0.1$) and across distinct regimes. In the TLL phase, the spin current follows $|\tilde{I}_{S,2}| \sim T^\alpha$, reflecting the gapless spinon excitation. Note that while such algebraic spinon spin current can be obtained in TLL theory with nonlinear spinon dispersion~\cite{Hirobe2017}, there are challenges to accurately determine the critical exponents $\alpha$~\cite{Supplementary}. At the QCP ($B_c=2$), we find a universal scaling $\tilde{I}_{S,2} \sim \sqrt{T}$ in both real- and imaginary-time approaches (see Appendix). These results demonstrate our approach as an accurate and efficient approach for SSE simulations. 

 {While the derivation of $\tilde{I}_2$ assumes a leading term $\mathrm{Im}[\chi_{\mathrm{loc}}^{-+}(\omega)] \sim f_2 \omega^2$, the approach is not restricted to this form~\cite{Supplementary}. The agreement between $\tilde{I}_S$ and $\tilde{I}_2$ indicates that Eq.~(\ref{Eq:I2}) remains valid even when $\mathrm{Im}[\chi_{\mathrm{loc}}^{-+}(\omega)]$ exhibits a fractional power-law dependence at low frequencies, as noted for 1D spin chain in Ref.~\cite{wang2025}.
Beyond bulk behavior, recent field-theoretical analyses of 1D TLLs predict a second sign reversal at lower temperatures under small magnetic fields $B \ll J$ once boundary contributions are included~\cite{wang2025}. Our calculations successfully reproduce this feature in Fig.~\ref{Fig2}(b), consistent with the theoretical prediction and potentially explaining earlier experimental results~\cite{Hirobe2017} (see Appendix).}

\textit{SSE in a triangular-lattice quantum antiferromagnet.---}
The easy-axis triangular-lattice antiferromagnet (TLAF) with $J_z > J_{xy}$ [Eq.~(\ref{Eq:H})] realizes the long-predicted spin supersolid state~\cite{Yamamoto2014,Sellmann2015,Gao2022Super}. This exotic phase has recently been experimentally observed in Co-based compounds \nbcp~\cite{Gao2022Super, Xiang2024Nature, Gao2024Dynam} and K$_2$Co(SeO$_3$)$_2$~\cite{zhu2024continuum, chen2024phase}. In the former, an effective model with coupling strength $J_{xy}=0.88~{\rm K}$, $J_z=1.48~{\rm K}$ accurately describes its magnetic properties~\cite{LiN2020, Lee2021, Xiang2024Nature} and spin dynamics~\cite{Wu2022PNAS, Gao2024Dynam, Sheng2025}. We hereafter simulate SSE in the easy-axis TLAF model using NBCP parameters, noting that our results also extend to other spin-supersolid materials like K$_2$Co(SeO$_3$)$_2$~with the similar model.

As observed in experiments~\cite{LiN2020,Wu2022PNAS,Xiang2024Nature} and comprehended in theoretical calculations~\cite{Gao2022Super}, NBCP exhibits four distinct phases: supersolid-Y (SSY), up-up-down (UUD), supersolid-V (SSV), and the polarized (PL) phases. They are separated by three QCPs located at $B_{c1}\simeq 0.35(5)$~T, $B_{c2} \simeq 1.15(4)$~T, and $B_{c3} \simeq 1.69(6)$~T~\cite{Xiang2024Nature}. In both SSY and SSV phases, the system exhibits simultaneous breaking of lattice translation and U(1) rotation symmetries, establishing a quantum magnetic analog of triangular-lattice supersolid~\cite{Melko2005, Wessel2005, Heidarian2005, Prokofev2005, WangF2009, Jiang2009}. 

Figure~\ref{Fig3}(a) reveals the simulated spin currents $\tilde{I}_2$, which can be used to map the phase diagram of NBCP. The different signs and temperature dependence of $\tilde{I}_2$ distinguish various spin states. Both supersolid phases (SSY and SSV) can be recognized by the negative spin currents, where the sign reversal marks the transition from higher-temperature states to the spin-supersolid phase. In contrast, in the UUD phase between $B_{c1}$ and $B_{c2}$, the spin current decays rapidly at low temperature [see Fig.~\ref{Fig3}(c)] due to its gapped nature; the PL regime shows persistently a positive sign.

Figure~\ref{Fig3}(b) demonstrates the precise detection of all three QCPs through SSE measurements. The peaks and dips in the spin current profile show excellent agreement with established QCP locations in prior studies~\cite{LiN2020, Gao2022Super, Xiang2024Nature}. Moreover, Fig.~\ref{Fig3}(c) shows the linear temperature dependence of $\tilde{I}_2$ near three QCPs, consistent with quantum critical scaling $\tilde{I}_2 \sim T^{d/z}$ ($d=2, z=2$) of Bose-Einstein condensation universality class~\cite{Giamarchi2008BEC, Zapf2014RMP}.
The universal spin currents are mediated by the gapless excitations at QCPs, reflecting the low-energy density of states encoded in the symmetric part of the local dynamical susceptibility $\frac{1}{2}{\rm Im}\left[\chi^{-+}_{\rm loc}(\omega) + \chi^{-+}_{\rm loc}(-\omega)\right] \sim \omega^{(d-z)/z}$ (see Appendix). Note  {such a universal} temperature scaling of spin current can also be captured by the spin-wave theory near the polarization field~\cite{Supplementary}.

\textit{Spin current sign reversal.---}
To understand the sign reversal in the spin supersolid phase, we decompose the local operator as $\mathcal{O}_j = \mathcal{O}_j^J + \mathcal{O}_j^B$, where $\mathcal{O}_j^J = [H_0, S_j^-] = \sum_{\langle i,j \rangle}(J_{xy}S_i^-S_j^z - J_zS_i^zS_j^-)$ and $\mathcal{O}_j^B = [-B\sum_i S_i^z, S_j^-] = BS_j^-$. We then compute the component $\tilde{I}_2^J = \beta \langle \mathcal{O}_j^J(\frac{\beta}{2}) S_j^+ \rangle_\beta $ from spin exchange and $\tilde{I}_2^B = \beta \langle \mathcal{O}_j^B(\frac{\beta}{2}) S_j^+ \rangle_\beta$ from the Zeeman coupling, with the total current $\tilde{I}_2 = \tilde{I}_2^J + \tilde{I}_2^B$. We find that the spin exchange generates a negative spin current ($\tilde{I}^J_2 < 0$) while the Zeeman term leads to positive contributions ($\tilde{I}^B_2 > 0$, see Appendix). Therefore, the sign reversal in spin supersolid phase can be regarded as a competition between exchange-coupling and Zeeman-term effects --- the interaction plays a dominant role at low temperatures and thus gives rise to a negative spin current. Note such sign reversal in spin supersolid is not captured by linear spin-wave theory~\cite{Supplementary}. Moreover, in the PL regime ($B \geq B_{c3}$), strong magnetic fields suppress exchange effects, resulting in exclusively positive spin currents across the whole temperature window. 

Within the UUD phase, we observe a field-driven sign reversal of the spin current --- positive at lower fields and negative at higher fields --- with the boundary at the 1/3-magnetization plateau midpoint [Fig.~\ref{Fig3}(a,b)]. This phenomenon can be explained by examining the temperature dependence of magnetization: At the plateau midpoint where $\frac{dM}{dT} = 0$, spin current vanishes when $M$ becomes temperature-independent. Moving away from this point, the sign of spin current follows $(- \frac{dM}{dT})$ --- positive for $\frac{dM}{dT} < 0$ and negative for $\frac{dM}{dT} > 0$. Since $\frac{dM}{dT}$ also quantifies the magnetocaloric effect (MCE), such observation reveals inherent connections between SSE and MCE~\cite{Supplementary}.

\begin{figure}[t]
\includegraphics[width=1\linewidth]{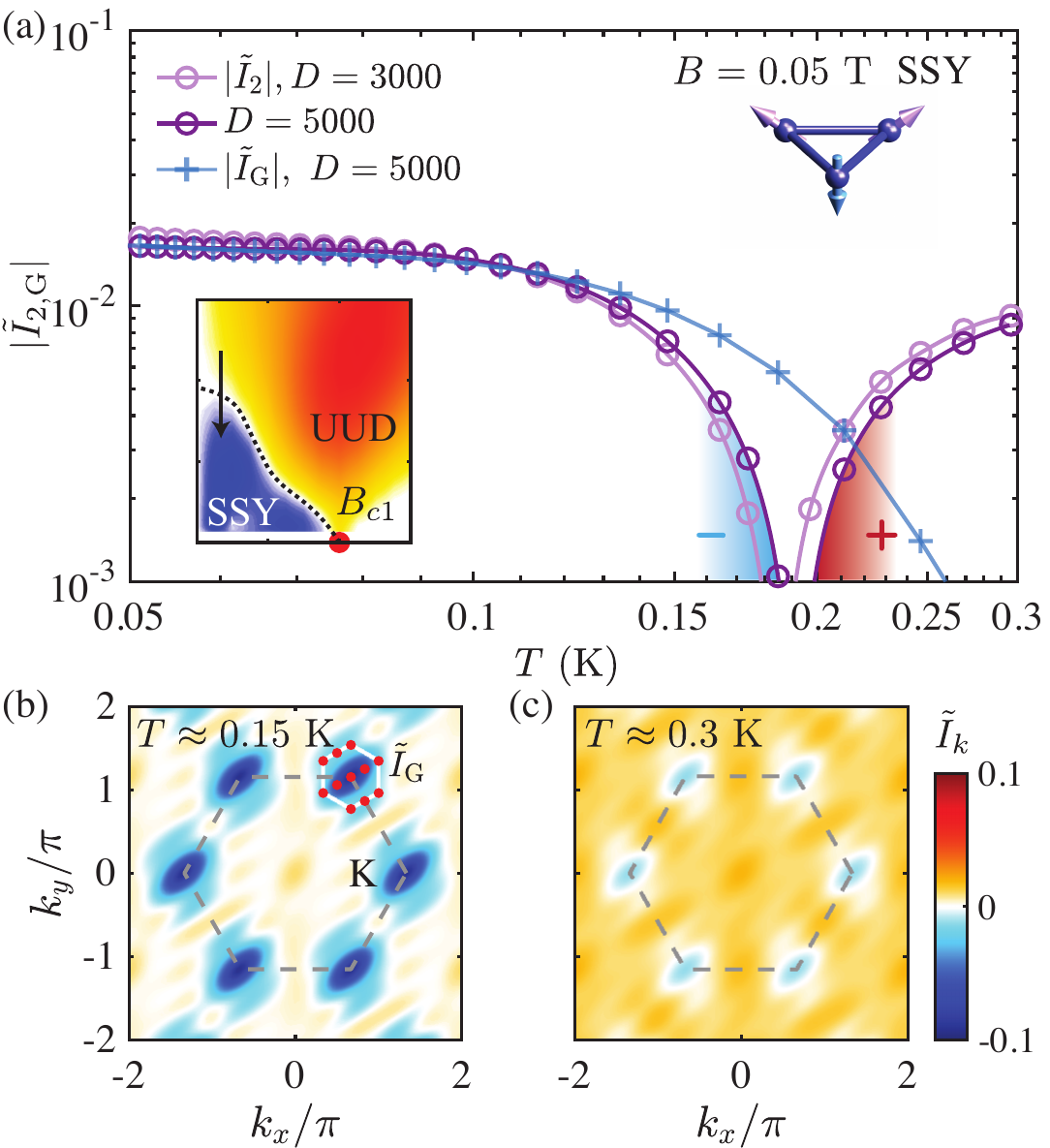}
\caption{(a) The simulated $\tilde{I}_2$ results in the SSY phase, where the data are well converged with $D=5000$. (b) and (c) present the momentum-resolved spin current $\tilde{I}_k$ at two temperatures. The gray dashed line shows the boundary of the 1st Brillouin zone. The red dots mark the involved momentum points in the calculations of $\tilde{I}_{\rm G}$. The black dots label the $\Gamma$ and K points.  {(a) shows the spin supercurrent under a 0.05~T field (see inset); results for a different field in SSY phase can be found in the Appendix.}
}
\label{Fig4}
\end{figure}

\textit{Spin supercurrent in the supersolid phase.---} 
The nonzero spin superfluid density --- unusual in easy-axis systems --- characterizes the spin supersolid phase, quantified by spin stiffness~\cite{Huang2025_Stiffness} and distinct transport signatures. Figure~\ref{Fig4}(a) reveals a striking spin current behavior across the UUD-to-SSY transition: Sign reversal upon entering the SSY phase, and persistent negative current that saturates to a nonzero value at low temperature, revealing a quantum transport signature of the spin supersolid phase. 
  
To elucidate the origin of negative spin supercurrents, we compute the momentum-resolved current $\tilde{I}_k = \beta \langle \mathcal{O}_{k}(\beta/2) S_k^+ \rangle$ (where $S_k^+ = \frac{1}{\sqrt{N}} \sum_n e^{ikn}S_n^+$ and $\mathcal{O}_k=[H, S_k^-]$), indicating distinct temperature-dependent behaviors in Fig.~\ref{Fig4}(b,c): At $T=0.3$ K, most momentum points contribute positively to the net current; while at $T=0.15$ K, gapless Goldstone modes near the K point dominate with negative contributions. By  {summing over} these modes through $\tilde{I}_{\rm G}=\frac{2}{N}\sum_{k\in k_{\rm G}}\tilde{I}_k$ [with $k_{\rm G}$ marked in Fig.~\ref{Fig4}(b)], we establish the saturated $\tilde{I}_{\rm G} \approx \tilde{I}_2$ at low temperatures [Fig.~\ref{Fig4}(a)]. Since Goldstone modes at the K-point carry positive angular momentum (in contrast to negative $\Gamma$-point modes), increasing K-magnon populations lead to $dM/dT > 0$ and consequently generate the negative spin supercurrents. The easy-axis system sustains persistent currents despite out-of-plane UUD ordering [Fig.~\ref{Fig4}(a) and inset] --- providing quantum transport probe of spin supersolidity with clear experimental signatures for future studies.  

{\textit{Discussion.---}}
We develop a thermal tensor-network approach to investigate SSE in quantum magnets. Our approach enables accurate calculations of spin currents and their temperature dependence,  {as benchmarked on 1D Heisenberg chain and applied to 2D frustrated triangular-lattice system. We uncover two sign reversals of the spin current in the 1D Heisenberg chain under a small magnetic field, consistent with previous field-theoretical predictions~\cite{wang2025}, and it may account for the experimental observations~\cite{Hirobe2017}.}
Remarkably, in triangular-lattice spin supersolid phase, we observe spin supercurrents that saturate at low temperatures --- a SSE signature directly linked to dissipationless Goldstone-mode excitations. Similar to the persistent residual angular momentum proposed for detecting the boson supersolid~\cite{Leggett1970}, it serves as  {a macroscopic quantum phenomenon} and sensitive indicator of spin supersolid. The $\tilde{I}_2$ results advance understanding of triangular-lattice spin supersolids and potentially opens new pathways for probing fractional-excitation spin currents in  {frustrated quantum spin systems, particularly the spin liquid systems}. Moreover, $\tilde{I}_2$ can be computed using a range of numerical approaches,  {including those based on} matrix-product states~\cite{White2009, Stoudenmire2010, Sugiura2013,Iwaki2021}, projected-entangled-pair operators~\cite{Li2011, Czarnik2012, Kshetrimayum2019Anneal, Czarnik2019,Wietek2019}, and quantum Monte Carlo samplings~\cite{Sandvik1991, Sandvik2010},  {demonstrating its broad applicability}. 

 {Regarding the SSE experiment,} while the supercurrent awaits  {to be observed} in the spin-supersolid materials like NBCP~\cite{Gao2022Super, Xiang2024Nature}, prior spin-current measurements in candidate spin-superfluid systems including FM film Y$_3$Fe$_5$O$_{12}$~\cite{Bozhko2016SF} and 3D compound Cr$_2$O$_3$~\cite{Yuan2018} demonstrate the experimental feasibility. Moreover, the inverse effect of SSE, the spin Peltier effect~\cite{Flipse2014PRL, Daimon2016, Ohnuma2017}, enables a new avenue for ultralow-temperature cooling  {with frustrated magnets}. Onsager reciprocity~\cite{Onsager1931} requires that the spin supersolids (and potentially other spin states) with strong SSE must also exhibit enhanced spin-current-driven cooling effect.  {Our work thus identifies spin current scaling as a sensitive probe of spin excitations in frustrated magnets, particularly in spin supersolids, positioning them as promising platforms for ultralow-temperature spin caloritronics.}

\begin{acknowledgements}
\textit{Acknowledgments.---}
W.L., and Y.G. are indebted to Ning Xi, Jianxin Gao, Enze Lv, Jiang Xiao, Oleg Starykh, Tao Shi, and Gang Su for insightful discussions. Y.H. express his gratitude to Masahiro Sato for stimulating discussions. This work was supported by the National Key Projects for Research and Development of China (Grant No.~2024YFA1409200), the National Natural Science Foundation of China (Grant Nos.~12222412 and 12447101), and Chinese Academy of Sciences under contract numbers XDB1270100 and YSBR-057. S.M. is supported by JSPS KAKENHI No. 24K00576 from MEXT, Japan. Y.G. and W.L. thank the HPC-ITP for the technical support and generous allocation of CPU time. The data that support the findings of this work are openly available~\cite{datalink}.
\end{acknowledgements}

\bibliography{SSERef.bib}
\onecolumngrid
\newpage
\begin{center}
\textbf{\large{Appendix}}
\end{center}
\twocolumngrid

\textit{Tensor network approach for spin current.---}
We employ thermal tensor-network approach to obtain the finite-temperature density matrix $\rho(\beta/2)$, with an efficient representation of matrix product operator (MPO)~\cite{tanTRG2023,Li2011,ChenSETTN2017,Chen2018}, enabling simulations of imaginary-time correlation functions. For the spin-1/2 Heisenberg chain, we perform calculations on  {both finite-size (up to $L=256$)~\cite{tanTRG2023} and infinite-size~\cite{Li2011} chains,} with retained bond dimension $D=500$. 
To benchmark the results, we also perform real-time evolution~\cite{TDVP2011,TDVP2016} on the density matrix MPO to compute the spin current $\tilde{I}_S$. In these real-time simulations, we maintain a bond dimension of $D = 500$ to ensure data convergence (see Supplementary Materials~\cite{Supplementary}).  {In addition}, we note that a recent work~\cite{wang2025} emphasizes boundary contributions  {are important for} spin currents in the 1D TLL chains,  {as will be discussed below}. In the simulations of the easy-axis TLAF model for NBCP, we map the system to a quasi-1D chain with long-range interactions~\cite{tanTRG2023,Chen2018}. Calculations are performed on a Y-type cylinder with width $W=6$ and length $L=18$~\cite{Supplementary}, with bond dimension up to $D=5000$. In practice, we compute bulk-averaged spin currents by excluding edge effects - specifically discarding three columns from both ends of the cylinder. 

\begin{figure}[]
\includegraphics[width=1\linewidth]{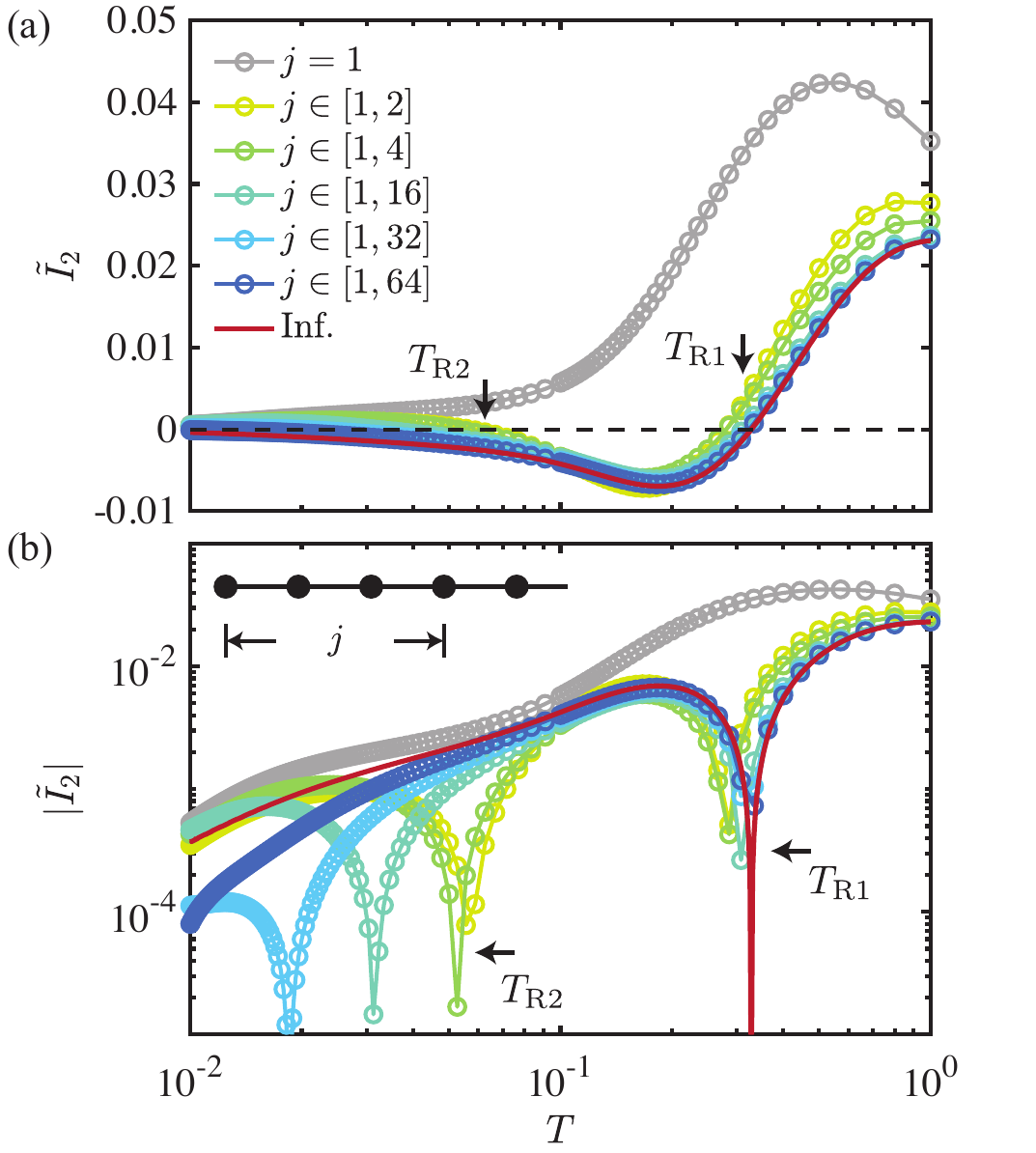}
\caption{ {The simulated spin current $\tilde{I}_2$ is shown both at the outermost site ($j=1$) and averaged over several adjacent edge sites ($j \in [1, l]$) on a spin-1/2 chain with length $L=256$. The bulk results computed on an infinite chain are also included. A magnetic field $B=0.1$ is applied and bond dimension $D=500$ is kept. The two arrows indicate the higher sign-reversal temperature $T_{\rm R1}$ and the lower temperature $T_{\rm R2}$, respectively.}}
\label{FigE1}
\end{figure}

 {
\textit{Boundary contributions in spinon spin current.---} For 1D TLL spin chain, boundary effects may play an essential role~\cite{wang2025} when comparing theory and experiments with a longitudinal SSE setup~\cite{Hirobe2017}. Therefore, we evaluate the spin current $\tilde{I}_2$ averaged over the first $l$ boundary sites ($j\in[1,l]$), as shown in Fig.~\ref{FigE1}. When only the outmost edge site is included ($j=1$), the spin current remains positive without any sign reversal, in contrast to experimental observations on spin-chain compound~\cite{Hirobe2017}. However, by including a few more sites (even for $l=2$) we observe two sign reversals in the spin current, one at higher temperature $T_{\rm R1} \simeq 0.3$ and the other at lower temperature $T_{\rm R2} \lesssim B$. The spin current is positive at the lowest temperatures, which is consistent with prior field-theoretical results~\cite{wang2025}. The results suggest that the low-temperature sign reversal reported in Ref.~\cite{Hirobe2017} may originate from boundary effects. The SSE length scale $l$ is governed by interface quality and electron penetration depth, calling for combined experimental and theoretical studies. 
From theoretical side, as $l$ increases, we find the lower sign-reversal temperature $T_{\rm R2}$ decreases. The results converge in the bulk limit to the infinite-chain scaling $\tilde{I}_2 \sim T^\alpha$, as obtained using the linearized tensor renormalization group approach~\cite{Li2011}.
Moreover, for the 2D triangular lattice, the results presented in Figs.~\ref{Fig3} and \ref{Fig4} --- including the sign reversal and saturated spin current at low temperature --- are robust and show minimal boundary effects (c.f. Fig.~\ref{FigE4}).}

\begin{figure}[t!]
\includegraphics[width=1\linewidth]{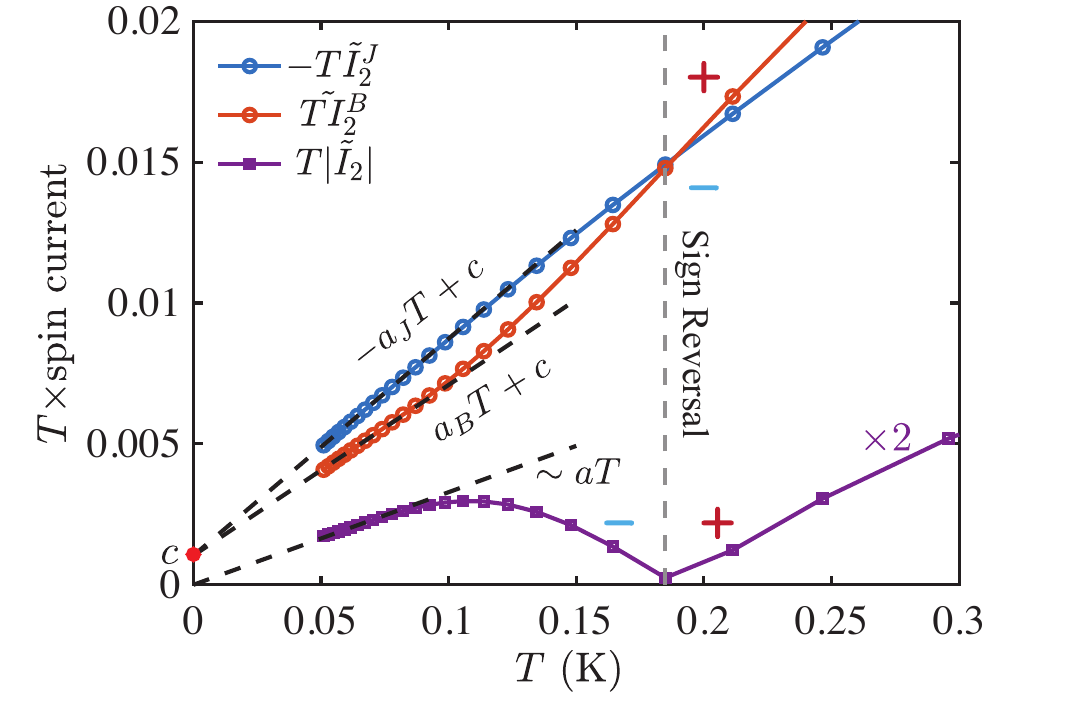}
\caption{Simulated spin current and its components $\tilde{I}_2=\tilde{I}_2^J+\tilde{I}_2^B$, computed in the SSY phase under a magnetic field $B=0.05~{\rm T}$  {(with coupling parameters for NBCP)}. The vertical gray dashed line indicates the location of sign reversal in net current $\tilde{I}_2$. The back dashed line shows the linear-fitting of the low-temperature $T\tilde{I}_2^{B,J} = a_{B,J} T \pm c$, with $a_B \simeq 0.061$, $a_J \simeq -0.077$, and $c\simeq 0.001$. The net spin current $T\tilde{I}_2$, values being amplified by twice in the plot, scales as $a \,T$ with $a = a_B + a_J \simeq 0.016$ at low temperature.
}
\label{FigE2}
\end{figure}

\textit{Sign reversal and spin supercurrent.---}
To understand the sign reversal of spin current, we decompose the total current as $\tilde{I}_2 = \tilde{I}_2^J + \tilde{I}_2^B$, with $\tilde{I}_2^J = \beta \langle \mathcal{O}_j^J (\beta/2) S_j^+\rangle$ and $\tilde{I}_2^B = \beta \langle \mathcal{O}_j^B (\beta/2) S_j^+\rangle$. Figure~\ref{FigE2} demonstrates a spin-current sign reversal in supersolid phase (SSY): $\tilde{I}_2^J$ (from spin interaction) is negative while $\tilde{I}_2^B$ (from Zeeman term) remains positive. The two contributions compete and cross at the sign-reversal temperature when the net current $\tilde{I}_2$ becomes negative. In addition, the observed sign reversal of spinon spin current in 1D Heisenberg chain (Fig.~\ref{Fig2}) can be explained in a similar way~\cite{Supplementary}. Furthermore, Fig.~\ref{FigE2} shows that the nonzero intercepts ($c$) of $T\tilde{I}_2^B$ and $T\tilde{I}_2^J$ cancel at $T=0$, as required by the ground-state identity $\langle[H, S_j^+]\rangle \equiv 0$. The persistent spin supercurrent $\tilde{I}_2 \sim a$ arises from the differing slopes $a_{B,J}$ of $T\tilde{I}_2^B$ and $-T\tilde{I}_2^J$, leading to a constant value $a = a_B + a_J$.

\textit{Derivation of spin-current universal scaling near QCP.---} 
Below we analyze the spin current at the spin polarization QCP ($B=B_c$) with U(1) symmetry, and the ground state becomes fully polarized for $B > B_c$. As the kernel function $k^2(\beta \omega)$ is an even function, we only consider the even part of ${\rm Im}[\chi^{-+}_{\rm loc} (\omega)]$, i.e., $X(\omega)\equiv \frac{1}{2}{\rm Im}[\chi^{-+}_{\rm loc} (\omega) + \chi^{-+}_{\rm loc} (-\omega)]$ with the corresponding spectral representations:
\begin{equation}
    \begin{split}
        X(\omega) =& \frac{\pi}{2\mathcal{Z}}\sum_{m,n}(|| \langle m| S_j^- |n \rangle ||^2 - || \langle m| S_j^+ |n \rangle ||^2) \\
        &\cdot e^{-\beta E_n} (1-e^{-\beta \omega}) \delta(\omega + E_n - E_m). 
    \end{split}
\end{equation}
In the low-temperature limit, we consider only the contributions from the positive energy part ($E_m > E_n$ and $\omega > 0$), 
\begin{equation}
X(\omega)=\frac{\pi}{2}\sum_{k}|| \langle k| S_j^- |{\rm PL} \rangle ||^2 \delta(\omega - \omega_{k}), 
\label{EqS:chi-+}
\end{equation}
where $|k \rangle = \frac{1}{\sqrt{N}} \sum_r e^{ikr}S^-_r | \rm{PL} \rangle$ is the single-magnon excited state with dispersion $\omega_k \sim (k-k_0)^z$, and $| \rm{PL} \rangle$ is the fully polarized state $\ket{\uparrow \uparrow \uparrow ... \uparrow}$. For the polarization QCP with U(1) symmetry, we have the dynamical exponent $z=2$.

\begin{figure}[htbp]
\includegraphics[width=1\linewidth]{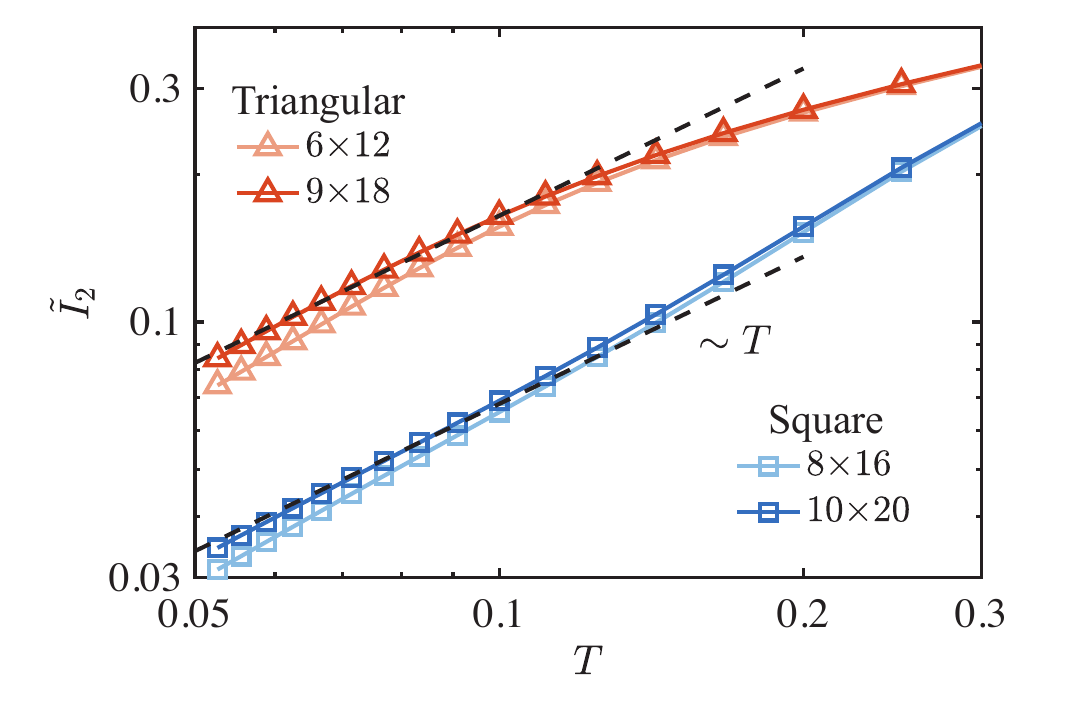}
\caption{Simulated spin current $\tilde{I}_2$ of square- and triangular-lattice Heisenberg models at the QCPs ($B_c=4$ and $B_c=4.5$, respectively). The calculations are conducted on the $W\times L$ cylinder and the retained bond dimension is $D=2000$.}
\label{FigE3}
\end{figure}
 
As $|| \langle k| S_j^- |0 \rangle ||^2 = ||\frac{1}{\sqrt N}e^{ikj}||^2 = \frac{1}{N}$ is a constant for any $k$, the quantity of interest, $X(\omega)$, can be represented as the density of states up to a constant. Based on Eq.~(\ref{EqS:chi-+}), we have $X(\omega) \sim \omega^{\frac{d-z}{z}}$ in the low-frequency regime, where $d$ is the dimension of the system. Substitute it into the expression of $\tilde{I}_S=\int_{-\infty}^\infty d\omega~k^2(\beta \omega)X(\omega)$, we arrive at $\tilde{I}_{S} \sim T^{d/z}$.
For 1D Heisenberg chain, this scaling reads $\tilde{I}_{S} \sim \sqrt{T}$, in consistent with the numerical results in Fig.~\ref{Fig2}(c). Beyond 1D chain, we further compute the spin current of 2D square- and triangular-lattice Heisenberg models at their polarized QCPs. As shown in Fig.~\ref{FigE3}, the low-temperature behavior exhibits a linear-$T$ scaling, i.e., $\tilde{I}_S \sim T$ ($d = z = 2$).  

{\textit{Extended spin-current data in spin supersolid phase.---}} 
In Fig.~\ref{Fig4}, we demonstrated the persistent spin supercurrents in the SSY phase at 0.05 T, exhibiting persistent supercurrent behavior mediated by the dissipationless Goldstone modes.   
Figure~\ref{FigE4} extends these observations to  {a different field ($B=0.2~{\rm T}$) in the SSY phase. In this case, we also find the spin current exhibits a sign reversal from positive values at high temperatures (UUD or paramagnetic phase) to negative values in the supersolid regime. Figure~\ref{FigE4} further shows $\tilde{I}_2$, averaged over the first $l$ columns ($j \in [1,l]$) from the cylinder boundary. Owing to the three-sublattice structure of the supersolid, the calculation begins from $l=3$. As $l$ increases from 3 to 9, the persistent sign reversal and low-temperature saturation of the spin supercurrent demonstrate that these phenomena are more robust against boundary effects in the 2D triangular lattice than in the 1D case. This firmly establishes the spin Seebeck response as an intrinsic hallmark of triangular spin supersolid phases.}

\begin{figure}[b]
\includegraphics[width=1\linewidth]{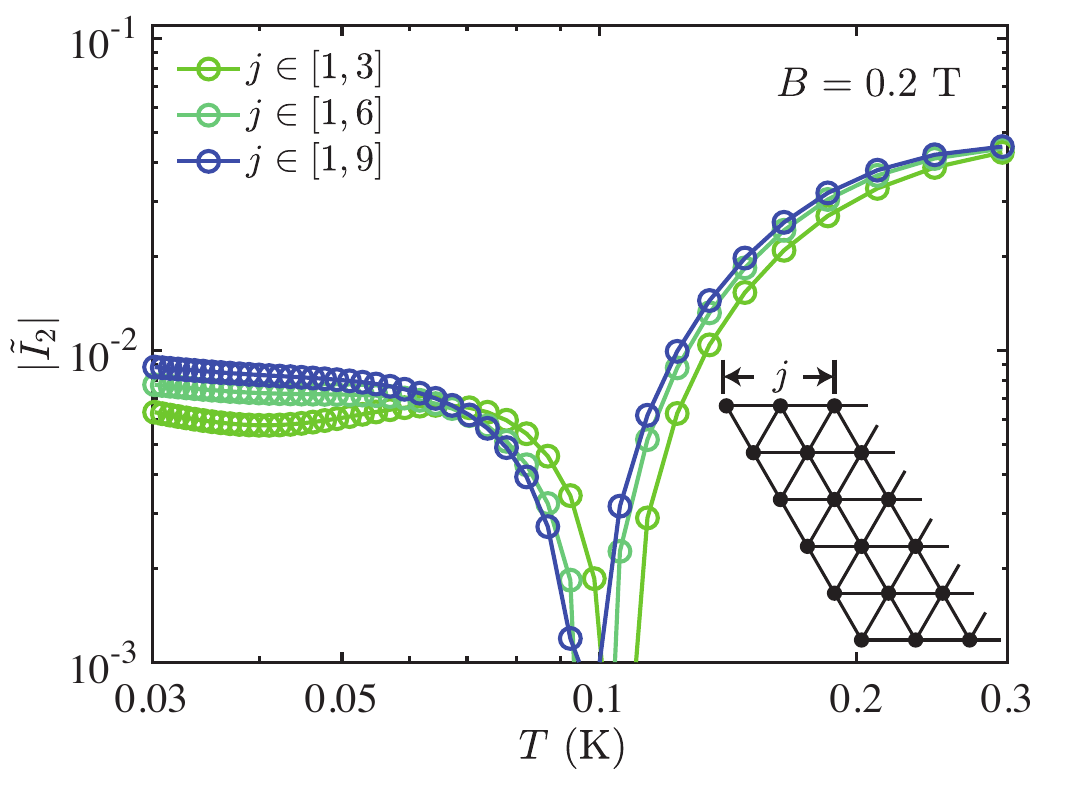}
\caption{ {The simulated spin current $\tilde{I}_2$ on a YC6$\times$18 lattice, averaged over the first $l$-column (denoted as $j \in [1, l]$). The simulations are conducted for the realistic spin model for \nbcp~under a small magnetic field $B=0.2~{\rm T}$. The retained bond dimension is $D=3000$, ensuring well converged results.}
}
\label{FigE4}
\end{figure}

%
\onecolumngrid
\newpage
\clearpage
\onecolumngrid
\mbox{}
\begin{center}
{\large Supplementary Materials for}
$\,$\\
\textbf{\large{Spin Seebeck Effect of Triangular-lattice Spin Supersolid}}

$\,$\\
Gao \textit{et al.}
\end{center}

\date{\today}
\setcounter{section}{0}
\setcounter{figure}{0}
\setcounter{equation}{0}
\setcounter{table}{0}
\renewcommand{\theequation}{S\arabic{equation}}
\renewcommand{\thefigure}{S\arabic{figure}}
\renewcommand{\thetable}{S\arabic{table}}
\setcounter{secnumdepth}{3}
\renewcommand{\theequation}{\Alph{section}\arabic{equation}}
\setcounter{page}{1}

\renewcommand\arraystretch{2}

\section{Derivation of the normalized spin current}
In this section, we show the detailed derivation of spin current in the spin Seebeck effect (SSE), i.e. Eq.~(1) in the main text~\cite{Adachi2011,Adachi2013,Masuda2024}. The full Hamiltonian describing the spin-metal junction can be expressed as
\begin{equation}
H = H_S + H_M + H_{\rm int}
\end{equation}
with 
\begin{equation}
\begin{split}
H_S &= \sum_{\langle i,j \rangle} \frac{J_{xy}}{2} (S_i^+S_j^- + S_i^-S_j^+) + J_z S_i^z S_j^z - B\sum_i S_i^z,\\
H_M &= \sum_{k, \sigma = \{\uparrow, \downarrow\}} 
\epsilon_{k,\sigma} f_{k, \sigma}^{\dagger} f_{k, \sigma},\\
H_{\rm int} &=  J_{sd} \sum_{i \in {\rm int}} {S}_i \cdot {s}_i,
\end{split}
\end{equation}
where $\sum_{i \in {\rm int}}$ stands for the summation over all the sites on the interface, $S$ is the spin operator of the insulator quantum magnet and $s$ is the electron spin operator in the metal side.  {We note that the spin–metal interface depends on the experimental setup, e.g., longitudinal~\cite{Hirobe2017} or transverse setup~\cite{Uchida2010Insulator}.}

The tunneling spin current is defined through the time derivative of the conduction electrons' spin-polarization density at the interface:
\begin{equation}
I_S = \sum_{i\in {\rm int}} \frac{\partial}{\partial t} s_i^z(t) = 
-i \sum_{i\in {\rm int}} [s_i^z(t), H] = J_{sd} \sum_{i\in {\rm int}} -i S_i^-(t) s_i^+(t) + h.c.
\end{equation}
The statistical average of $I_S$ under the non-equilibrium steady state in the SSE experimental setup is given by
\begin{equation}
\langle I_S \rangle = 2 J_{sd} \sum_{i\in {\rm int}} {\rm Re} [(-i) \langle S_i^-(t) s_i^+(t) \rangle]. 
\label{Eq:SMsd}
\end{equation}

Given the relative weakness of the $s-d$ coupling compared to the energy scales of both the metal and magnet,
we treat $H_S$ and $H_M$ as unperturbed Hamiltonian while considering $H_{\rm int}$ as a perturbation.
Assuming randomly distributed interface sites with inter-site distances significantly exceeding
the lattice constants of both magnet and metal, we derive:
\begin{equation}
\langle I_S \rangle =2 N_{\rm int} J_{sd} \lim_{\delta \rightarrow 0^+} {\rm Re} [F_{+-}^{<}(t, t' = t+\delta)],
\label{EqIS1}
\end{equation}
where $F_{+-}^{<}(t, t') = -i \langle S_i^{-}(t) s_i^+(t')\rangle$ 
and $N_{\rm int}$ is the number of the interaction sites.

Expanding the exponential factor in the statistical average of
$F_{+-}(t,t') = -i \langle T_C s_i^+(t) S_i^- (t) \rangle$ with respect to $H_{\rm int}$:
\begin{equation}
\begin{split}
F_{+-}(t,t') &= -i \sum_{n=0}^{\infty} \frac{(-i)^n}{n!} \int_C d t_1 \cdot\cdot\cdot \int_C d t_n 
\langle T_C \tilde{s}_i^+(t) \tilde{S}_i^-(t') \tilde{H}_{\rm int}(t_1) \cdot\cdot\cdot \tilde{H}_{\rm int}(t_n) \rangle_0 \\
&= (-i)^2 \int_C d t_1 \langle T_C \tilde{s}_i^+(t) \tilde{S}_i^-(t') \tilde{H}_{\rm int}(t_1)\rangle_0 + \cdot\cdot\cdot,
\end{split}
\end{equation}
where $\tilde{}$ stands for the time evolution under the unperturbed Hamiltonian, $\langle \cdot \cdot \cdot \rangle_0$ 
stands for the statistical average of the unperturbed Hamiltonian and $T_C$ is the time-ordered product on the Keldysh contour. 
Since the perturbed Hamiltonian is given by
\begin{equation}
\tilde{H}_{\rm int}(t_1) = J_{sd} \sum_{i \in {\rm int}} \tilde{ S}_i (t_1) \cdot \tilde{s}_i (t_1),
\end{equation}
we have 
\begin{equation}
\begin{split}
F_{+-}(t,t') &= J_{sd} \frac{(-i)^2}{2} \int_C d t_1 \langle T_C \tilde{s}_i^+(t) \tilde{s}_i^-(t_1) \rangle_0
\langle T_C \tilde{S}_i^+(t_1) \tilde{S}_i^-(t') \rangle_0 \\
&= \frac{J_{sd}}{2} \int_C d t_1~ X_{+-}(t, t_1) \chi_{+-}(t_1, t'),
\end{split}
\end{equation}
where
\begin{equation}
\begin{split}
X_{+-}(t, t') &= -i \langle T_C \tilde{s}_i^+(t) \tilde{s}_i^-(t') \rangle_0 \\
\chi_{+-}(t,t') &= -i \langle T_C \tilde{S}_i^+(t) \tilde{S}_i^-(t') \rangle_0.
\end{split}
\end{equation}
Using the Langreth rule, we have
\begin{equation}
F_{+-}^<(t,t') =\frac{J_{sd}}{2}  \int_{-\infty}^{\infty} d t_1 [X_{+-}^{\rm R}(t,t_1)\chi_{+-}^<(t_1, t') + X_{+-}^<(t,t_1)\chi_{+-}^A(t_1, t')],
\end{equation}
with 
\begin{equation}
\begin{split}
X_{+-}^{\rm R}(t) &= -i \theta(t) \langle [\tilde{s}_i^+(t), {s}_i^-]\rangle_0, \\
X_{+-}^<(t) &= -i \langle \tilde{s}_i^- (t) \tilde{s}_i^+\rangle_0, \\
\chi_{+-}^A(t) &= i \theta(-t) \langle [\tilde{S}_i^+(t), \tilde{S}_i^-]\rangle_0, \\
\chi_{+-}^<(t) &= -i \langle \tilde{S}_i^-(t) \tilde{S}_i^+\rangle_0. 
\end{split}
\end{equation}

Finally, applying the Fourier transformations, we arrive at
\begin{equation}
F_{+-}^<(t,t') =\frac{J_{sd}}{4 \pi} \int_{-\infty}^{\infty} d \omega~e^{-i\omega(t-t')} 
 [X_{+-}^{\rm R}(\omega)\chi_{+-}^<(\omega) + X_{+-}^<(\omega)\chi_{+-}^A(\omega)].
\label{EqGless}
\end{equation}
Put Eq.~(\ref{EqGless}) into Eq.~(\ref{EqIS1}), we have
\begin{equation}
\langle I_S \rangle = \frac{N_{\rm int} J_{sd}^2}{2 \pi} \int_{-\infty}^{\infty} d \omega
~ {\rm Re}[X_{+-}^{\rm R}(\omega)\chi_{+-}^<(\omega) + X_{+-}^<(\omega)\chi_{+-}^A(\omega)].
\end{equation}

Considering the following relationship:
\begin{equation}
\begin{split}
G^< (\omega) &= 2i {\rm Im}[G^{\rm R}(\omega)] n(T), \\
G^A(\omega) &= G^{\rm R}(\omega)^*,\\
n(T) &= \frac{1}{e^{\omega/T} - 1},
\end{split}
\end{equation}
we have
\begin{equation}
\begin{split}
\langle I_S \rangle =& \frac{N_{\rm int} J_{sd}^2}{2 \pi} \int_{-\infty}^{\infty} d \omega
~{\rm Re}[2i X_{+-}^{\rm R}(\omega) {\rm Im}[\chi_{+-}^{\rm R}(\omega)] n(T_s) + 
2i {\rm Im}[X_{+-}^{\rm R}(\omega)] n(T_m)  \chi_{+-}^{\rm R}(\omega)^* ]\\
=&  \frac{N_{\rm int} J_{sd}^2}{\pi} \int_{-\infty}^{\infty} d \omega~
-{\rm Im}[X_{+-}^{\rm R}(\omega)] {\rm Im}[\chi_{+-}^{\rm R}(\omega)] n(T_s)
+{\rm Im}[X_{+-}^{\rm R}(\omega)] {\rm Im}[\chi_{+-}^{\rm R}(\omega)] n(T_m)\\
=& \frac{N_{\rm int} J_{sd}^2}{\pi} \int_{-\infty}^{\infty} d \omega~
{\rm Im}[X_{+-}^{\rm R}(\omega)] {\rm Im}[\chi_{+-}^{\rm R}(\omega)] (n(T_m) - n(T_s)).
\end{split}
\end{equation}
We adopt the following approximations:
\begin{equation}
\begin{split}
{\rm Im}[X_{+-}^{\rm R}(\omega)]  \simeq & -a^2 \omega,\\
n(T_s) - n(T_m) \simeq&  \frac{\omega \delta T}{4 T^2 \sinh^2(\omega/(2T))},
\end{split}
\end{equation}
where $a^2$ is a constant, $\delta T = T_s - T_m$ and $T=(T_s + T_m)/2$. Note 
that $\chi_{+-}^{\rm R}(\omega) = -\chi_{-+}^{\rm R}(-\omega)$, we have
\begin{equation}
\begin{split}
\langle I_S \rangle =&  \frac{N_{\rm int} J_{sd}^2 a^2 \delta T}{4\pi T^2} \int_{-\infty}^{\infty} d \omega~
{\rm Im}[\chi_{+-}^{\rm R}(\omega)] \frac{\omega^2}{\sinh^2(\beta \omega/2)}\\
=&  -\frac{N_{\rm int} J_{sd}^2 a^2 \delta T}{4\pi} \int_{-\infty}^{\infty} d \omega~
{\rm Im}[\chi_{-+}^{\rm R}(\omega)] \frac{(\beta\omega)^2}{\sinh^2(\beta\omega/2)}\\
=& -A \delta T \tilde{I}_S,
\end{split}
\end{equation}
where $A = \frac{1}{4\pi} N_{\rm int} J_{sd}^2 a^2$ represents a material-dependent constant, 
$\delta T$ denotes the temperature gradient, $\beta \equiv 1/T$ is the inverse temperature, and 
the normalized spin current $\tilde{I}_S$ emerges as:
\begin{equation}
\tilde{I}_S =  \int_{-\infty}^{\infty} d \omega~{\rm Im}[\chi_{-+}^{\rm R}(\omega)] 
\frac{(\beta\omega)^2}{\sinh^2(\beta\omega/2)}.
\end{equation}

Our present theoretical framework focuses on the intrinsic bulk properties through simulations of dynamical susceptibility and spin currents, aligning with Refs.~\cite{Hirobe2017, Masuda2024, Kato2025PRX}. In realistic setup, there are additional complexities due to interfacial disorder, electron tunneling effects, and edge contribution~\cite{wang2025} --- all of which must be properly accounted for when comparing with experiments.

\section{Imaginary time approximation for spin current}

\subsection{ {General framework of imaginary-time approximation}}
In this section, we present detailed derivation of the imaginary time approximation for the SSE. The spin current $I_S = -A \tilde{I}_S \delta T$ is induced by both the magnetic field and temperature gradient, where the normalized spin current 
$\tilde{I}_S$ is given by
\begin{equation}
\tilde{I}_S = \int_{-\infty}^{\infty} d\omega~k^2(\beta \omega) 
{\rm Im}[\chi^{-+}_{\rm loc} (\omega)],
\end{equation}
with the dynamical susceptibility (retarded Green's function)
\begin{equation}
    \chi^{-+}_\mathrm{loc}(\omega) \equiv \chi_{-+}^{\rm R}(\omega) = -i \int_0^\infty ~ dt ~ 
    \langle[S_j^-(t), S_j^+]\rangle_T e^{i\omega t},
    \label{Eq:Chi}
\end{equation}
and $k^2(x\equiv \beta\omega )=x^2/\sinh^2{(x/2)}$.
We assume that ${\rm Im}[\chi^{-+}_{\rm loc} (\omega)]$ is analytical 
near $\omega = 0$, i.e.
\begin{equation}
{\rm Im}[\chi^{-+}_{\rm loc} (\omega)] = \sum_{n=1} \frac{\omega^n}{n!} f_n.
\label{EqS:Expand}
\end{equation}
Since the integral kernel $k(x\equiv \beta \omega) = \frac{x}{\sinh(x/2)}$ 
is an even function of $\omega$, only the even terms in 
Eq.~(\ref{EqS:Expand}) contribute. Given 
${\rm Im}[\chi^{-+}_{\rm loc} (0)]=0$, we obtain
\begin{equation}
    \begin{split}
    \tilde{I}_S &= \sum_{n=1} \int_{-\infty}^{\infty} 
    d\omega~k^2(\beta \omega) \frac{\omega^{2n}}{(2n)!} f_{2n} \\
    &= \sum_{n=1} \frac{1}{\beta^{2n+1}} 
    \int_{-\infty}^{\infty} dx~k^2(x) 
    \frac{x^{2n}}{(2n)!} f_{2n}\\
    &= \sum_{n=1} \frac{F_{2n}}{\beta^{2n+1}} f_{2n} \\
    &= \frac{16 \pi^4}{15 \beta^3} f_2 + O(\frac{1}{\beta^5}),
    \end{split}
    \label{EqS:Is}
\end{equation}
where $F_n \equiv \int_{-\infty}^{\infty} dx~k^2(x) \frac{x^{n}}{n!}$.

Considering the relationship between the imaginary-time correlation 
function and dynamical susceptibility, i.e.
\begin{equation}
    \langle S_j^-(\tau) S_j^+ \rangle = -\frac{1}{\pi} \int_{-\infty}^{\infty} d\omega~ \frac{e^{-\tau \omega}}
{1-e^{-\beta \omega}} {\rm Im} [\chi_{\rm loc}^{-+} (\omega)], 
\end{equation}
we have
\begin{equation}
    \frac{\partial}{\partial \tau} \langle S_j^-(\tau) S_j^+ \rangle = 
    \frac{1}{\pi} \int_{-\infty}^{\infty} d\omega~ \frac{\omega e^{-\tau \omega}}
{1-e^{-\beta \omega}} {\rm Im} [\chi_{\rm loc}^{-+} (\omega)].
\end{equation}
Given $\tau = \beta/2$, we have 
\begin{equation}
\begin{split}
    \frac{\partial}{\partial \tau}\langle S^-_j(\tau) S^+_j \rangle|_{\tau = \beta/2} &=
    \frac{1}{\pi} \int_{-\infty}^{\infty} d\omega~ \frac{\omega e^{- \beta \omega/2}}
    {1-e^{-\beta \omega}} {\rm Im} [\chi_{\rm loc}^{-+}(\omega)] \\
    &= \frac{1}{2\beta \pi} \int_{-\infty}^{\infty} d\omega~ k(\beta \omega) 
    {\rm Im} [\chi_{\rm loc}^{-+}(\omega)]\\
    &= \frac{1}{2 \beta \pi} \sum_{n=1} \int_{-\infty}^{\infty} d\omega~ 
    k(\beta \omega) \frac{\omega^{2n}}{(2n)!} f_{2n} \\
    &= \frac{1}{2 \beta \pi} \sum_{n=1} \frac{1}{\beta^{2n + 1}}
    \int_{-\infty}^{\infty} dx~ 
    k(x) \frac{x^{2n}}{(2n)!} f_{2n} \\
    &= \frac{1}{2 \beta \pi} \sum_{n=1} \frac{G_{2n}}{\beta^{2n + 1}} f_{2n}\\
    &= \frac{\pi^3}{\beta^4}f_2 + O(\frac{1}{\beta^6}),
\end{split}
\label{EqS:Corr}
\end{equation}
where $G_n \equiv \int_{-\infty}^{\infty} dx~k(x) \frac{x^{n}}{n!}$.

By comparing Eq.~(\ref{EqS:Is}) with Eq.~(\ref{EqS:Corr}), we have
\begin{equation}
    \tilde{I}_S = \frac{16 \pi \beta}{15}
    \frac{\partial}{\partial \tau}\langle S_j^-(\tau) S_j^+ \rangle|_{\tau = \beta/2} 
    + O(\frac{1}{\beta^5})
\end{equation}
Thus at low temperature, we obtain the imaginary time approximation 
$\tilde{I}_2$ of the normalized spin current $\tilde{I}_S$ following as
\begin{equation}
    \tilde{I}_S \sim \beta \frac{\partial}{\partial \tau}\langle S_j^-(\tau) S_j^+ \rangle|_{\tau = \beta/2}.
\end{equation}
The correlation function derivation can be implemented through the following steps
\begin{equation}
\begin{split}
\frac{\partial}{\partial \tau}\langle S_j^-(\tau) S_j^+ \rangle =&
\frac{1}{Z} {\rm Tr}[e^{-\beta H}e^{\tau H}HS_j^-e^{-\tau H}S_j^+ - 
e^{-\beta H}e^{\tau H}S_j^-He^{-\tau H}S_j^+]\\
=&\frac{1}{Z}{\rm Tr} [e^{-\beta H}e^{\tau H}[H, S_j^-]e^{-\tau H}S_j^+]\\
=&\langle \mathcal{O}_j(\tau) S_j^+\rangle,
\end{split}
\end{equation}
with $\mathcal{O}_j = [H, S_j^-]$. Finally, we arrive at  {the imaginary-time approximation of $\tilde{I}_S$}
\begin{equation}
\tilde{I}_2 \equiv \beta \langle \mathcal{O}_j(\frac{\beta}{2}) S_j^+\rangle \,  {\sim \tilde{I}_S}.
\end{equation}

\subsection{ {Extension to fractional power of $\omega$}}
 { This framework can be straightforwardly extended to cases where the leading term is fractional in $\omega$. For instance, in 1D spin chains the leading contribution to ${\rm Im} \chi_{\rm loc}^{-+}$ scales as $\omega^\alpha$ with non-integer $\alpha$, which does not alter the subsequent derivation. Specifically, the dynamical spin susceptibility is assumed to take the form
\[
{\rm Im}[\chi_{\rm loc}^{-+} (\omega)] = f_\alpha \, {\rm Re}[\omega^\alpha] + O(\omega),
\]
with a coefficient $f_\alpha$. The higher-order term $O(\omega)$ can be neglected at low temperatures, and we obtain
\begin{equation}
\begin{split}
    \tilde{I}_S &= \int_{-\infty}^\infty d\omega \, k^2(\beta\omega) \, \mathrm{Im}\!\left[\chi^{-+}_\mathrm{loc}(\omega)\right] \\
    &\simeq f_\alpha \, {\rm Re}\!\left[\int_{-\infty}^\infty d\omega \, k^2(\beta\omega) \, \omega^\alpha\right] \\
    &\simeq \frac{f_\alpha}{\beta^{1+\alpha}} \, {\rm Re}\!\left[\int_{-\infty}^\infty dx \, k^2(x) \, x^\alpha\right],
\end{split}
\end{equation}
and
\begin{equation}
\begin{split}
    \tilde{I}_2 = \beta \frac{\partial}{\partial \tau} \langle S^-_j(\tau) S^+_j \rangle \Big|_{\tau = \beta/2}
    & = \frac{1}{2 \pi} \int_{-\infty}^\infty d\omega \, k(\beta\omega) \, \mathrm{Im}\!\left[\chi_{\rm loc}^{-+}(\omega)\right] \\
    &\simeq \frac{f_\alpha}{2 \pi} \, {\rm Re}\!\left[\int_{-\infty}^\infty d\omega \, k(\beta\omega) \, \omega^\alpha\right] \\
    &\simeq \frac{f_\alpha}{2 \pi \beta^{1+\alpha}} \, {\rm Re}\!\left[\int_{-\infty}^\infty dx \, k(x) \, x^\alpha\right].
\end{split}
\end{equation}

Therefore, both $\tilde{I}_S$ and $\tilde{I}_2$ scale the same as $\frac{1}{\beta^{(1+\alpha)}}$, up to different prefactors, and share the same sign determined by the coefficient $f_\alpha$. This demonstrates that the derivation of the $\tilde{I}_2$ expression is general and also applies to the spin-1/2 chain case, which can more generally be denoted as $\tilde{I}_\alpha$.
}

\begin{figure*}[htpb]
\includegraphics[width=1\linewidth]{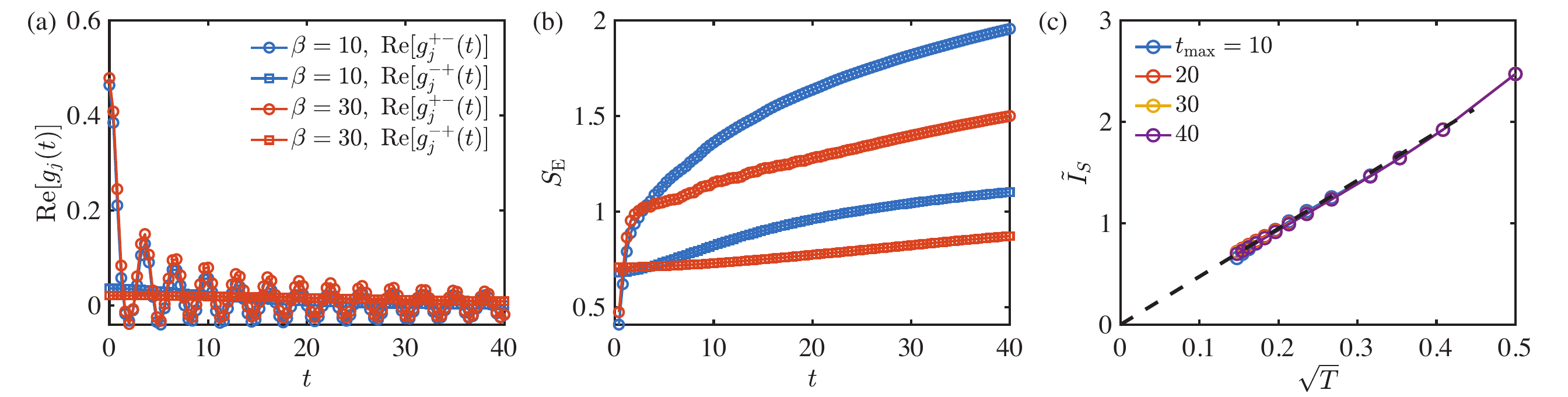}
\caption{(a) Real part of the real-time Green's function at different temperatures under a field of $B=B_c$ (QCP). (b) The entanglement entropy of the time-evolved state $\tilde{\rho}(t)$, sharing the same legend in (a). (c) The normalized spin current computed with different $t_{\rm max}$. In practical simulations, we evaluate the local Green's function at the central site $j=64$ of $L=128$ chain, with retained bond dimension $D=500$.}
\label{FigS1}
\end{figure*}

\section{Tensor-network approach for  {renormalized spin current from zero- and} finite-temperature spin dynamics}
\subsection{ {Spin current calculations from real-time dynamics at finite temperature}}
To compute the normalized spin current in Eq.~(2) of the main text, we evaluate the finite-temperature local dynamical susceptibility (Eq.~(\ref{Eq:Chi})) using the real-time Green's functions  {$g_j^{-+}(t, T)$} and  {$g_j^{+-}(t,T)$}, defined as follows:
\begin{equation}
\begin{split}
     {g_j^{-+}(t,T)} &\equiv \langle S_j^-(t) S_j^+ \rangle_T 
    = \frac{1}{\mathcal{Z}} {\rm Tr}[e^{-\beta H} e^{iHt}S_j^- e^{-iHt} S_j^+] \\
     {g_j^{+-}(t,T)} &\equiv \langle S_j^+(t) S_j^- \rangle_T
    = \frac{1}{\mathcal{Z}} {\rm Tr}[e^{-\beta H} e^{iHt}S_j^+ e^{-iHt} S_j^-]. \\
\end{split}
\end{equation}
Substituting them into Eq.~(\ref{Eq:Chi}), the local susceptibility reads  {(omitting the temperature index below)}:
\begin{equation}
\begin{split}
    \chi_{\rm loc}^{-+} (\omega) =& -i \int_0^{\infty} dt~ e^{i \omega t} (g_j^{-+}(t)-g_j^{+-}(-t)) \\
    =&  \int_0^{\infty} dt~ e^{i \omega t}
    ({\rm Re}[g_j^{-+}(t)] + i{\rm Im}[g_j^{-+}(t)] - {\rm Re}[g_j^{+-}(t)] + i{\rm Im}[g_j^{+-}(t)]) \\
    =& \int_0^{\infty} dt~ \sin(\omega t) ({\rm Re}[g_j^{-+}(t)] - {\rm Re}[g_j^{+-}(t)]) 
    + \cos(\omega t) ({\rm Im}[g_j^{-+}(t)] + {\rm Im}[g_j^{+-}(t)]) \\
    &+ i \int_0^{\infty} dt~ \cos(\omega t)({\rm Re}[g_j^{+-}(t)] - {\rm Re}[g_j^{-+}(t)]) 
    + \sin(\omega t)({\rm Im}[g_j^{-+}(t)] + {\rm Im}[g_j^{+-}(t)])
\end{split}
\end{equation}
Noting that the kernel function is even in $\omega$, we retain only the even part of 
$\chi_{\mathrm{loc}}^{-+} (\omega)$, leading to:
\begin{equation}
\tilde{I}_S = 2 \int_0^{\infty} d \omega~k^2(\beta \omega) \int_0^\infty dt~\cos(\omega t) 
{\rm Re}[g_j^{+-}(t)-g_j^{-+}(t)],
\label{Eq:G2SC}
\end{equation}
with which the normalized spin current $\tilde{I}_S$ can be obtained by computing the real-time correlation ${\rm Re}[g_j^{+-}(t)-g_j^{-+}(t)]$.

We calculate the real-time correlation functions through three major steps:
\begin{itemize}
\item[1.] Construct the finite-temperature density matrix $\rho(\beta/2) = e^{-\beta H/2}$ using tanTRG~\cite{tanTRG2023};
\item[2.] Compute the time-evolved state $\tilde{\rho}(t) = e^{-iHt} S_j^+ \rho(\beta/2) e^{iHt}$ via time-dependent variational principle (TDVP)~\cite{TDVP2011,TDVP2016};
\item[3.] Evaluate the Green function $g_j^{-+}(t) = \frac{1}{\mathcal{Z}}{\rm Tr}[{\rho^\dagger (\beta/2) \tilde{\rho}(t)}]$ at each time step.
\end{itemize}
For the second step, while the original TDVP algorithm was formulated for matrix product states, it can be naturally generalized to MPO --- see Ref.~\cite{tanTRG2023} for a concrete implementation.

Figure~\ref{FigS1}(a) displays the real-time Green's functions of the 1D Heisenberg model simulated at the polarization QCP ($B_c=2$). The real component of $g_j^{+-}(t)$ exhibits significantly greater magnitude than that of $g_j^{-+}(t)$, with this disparity becoming increasingly pronounced at lower temperatures. This behavior is consistent with ground-state property, where $g_j^{-+}(t)$ strictly vanishes in the fully polarized state.

Figure~\ref{FigS1}(b) shows the time evolution of the purified entanglement entropy $S_E$. While $S_E$ grows during time evolution, a bond dimension of $D = 500$ remains sufficiently large ($e^{\max(S_E)} \approx 7.3891 \ll 500$). This stands in sharp contrast to 2D systems, where the entanglement entropy of $\rho(\beta/2)$ exhibits extensive scaling, making finite-temperature real-time evolution computationally intractable.

\begin{table}[t]
\centering
\begin{tabular}{c|c|c|c}
\hline\hline
 & $\tilde{I}_S$ & $\tilde{I}_{S,{\rm G}}$ & $\tilde{I}_2$ \\
\hline
$T$-dependence 
& $k^2(\beta\omega)  {\rm Im}[\chi_{\rm loc}^{-+}(\omega)]$ 
& $k^2(\beta \omega)$
& $k(\beta\omega)  {\rm Im}[\chi_{\rm loc}^{-+}(\omega)]$ \\
\hline
Cost 
& \makecell{High}
& \makecell{High} 
& \makecell{Moderate} \\
\hline
Time evolution 
& Real and imaginary time
& Real time
& Imaginary time\\
\hline
{Value}
& {Complex number}
& {Complex number}
& {Real number} \\
\hline
\makecell{Steps}
& \makecell{$\tilde{N}_{\beta}N_{t}$ \\ $\sim 4000$ (spin chain)}
& \makecell{$N_{t}$ \\ $\sim 400$ (spin chain) }
& \makecell{$N_{\beta}$ \\ $\sim 100$ (spin chain)} \\
\hline\hline
\end{tabular}
\caption{ {Summary of three methods for evaluating the spin current. The calculations of $\tilde{I}_{S}$ and $\tilde{I}_{S,G}$ involve complex-number real-time evolution, which incurs a high computational cost. In contrast, $\tilde{I}_2$ requires only real-number calculations, resulting in significantly lower computational demands. $N_\beta$ is the number of imaginary-time evolutions, $N_t$ the number of real-time evolutions, and $\tilde{N}_\beta \lesssim N_\beta$ is the number of temperature points selected for successive real-time calculations. The listed values are typical parameters used in the 1D Heisenberg chain calculations.}}
\label{tab:Comp}
\end{table}

Figure~\ref{FigS1}(c) demonstrates improved low-temperature scaling with increasing $t_{\mathrm{max}}$, which is introduced in the computation of normalized spin current, i.e.,
\begin{equation}
    \tilde{I}_S \simeq 2 \int_0^{\infty} d \omega~k^2(\beta \omega) \int_0^{t_{\rm max}} dt~\cos(\omega t) 
    {\rm Re}[g_j^{+-}(t)-g_j^{-+}(t)]. 
\end{equation}
As the kernel function's $k^2(\beta\omega)$ emphasis on low-frequency components at low temperatures, longer evolution time $t_{\mathrm{max}}$ is needed to capture the dominant low-frequency dynamics.

As a sanity check, we verify the accuracy of our real-time evolution by numerically comparing both sides of the equation
\begin{equation}
 {\tilde{I}_2 \equiv } \beta\frac{\partial}{\partial \tau} \langle S^-_j(\tau) S^+_j \rangle \Big|_{\tau = \beta/2} = \frac{1}{2 \pi} \int_{-\infty}^\infty d\omega \, k(\beta\omega) \, {\rm Im} [\chi_{\rm loc}^{-+}(\omega)],
\end{equation}
with $j=64$ at the center of the chain. In practice, we find the relative difference is below $4\times10^{-4}$, indicating a well-converged real-time dynamical calculations with bond dimension $D=500$.

\subsection{ {Spin current calculations from zero-temperature dynamics}}
 {An alternative approach is to compute the spin current from ground-state spin dynamics, incorporating the temperature dependence solely through the kernel function~\cite{Kato2025PRX}. To be specific, we have 
\begin{equation}
    \tilde{I}_{S, {\rm G}} \simeq 2 \int_0^{\infty} d \omega~k^2(\beta \omega) \int_0^{t_{\rm max}} dt~\cos(\omega t) 
    {\rm Re}[g_j^{+-}(t,T=0)-g_j^{-+}(t, T=0)], 
\end{equation}
where $g^{ab}_j(t,T=0)=\langle\psi| e^{iHt} S_j^a e^{-iHt}S_j^b |\psi\rangle$ is the ground-state correlation function with $|\psi\rangle$ the ground state.

As shown in Fig.~\ref{FigS2}(a), $\tilde{I}_{S,{\rm G}}$ also exhibits sign reversals, albeit at higher temperatures than the accurate $\tilde{I}_S$. In the intermediate-temperature regime, it exhibits an algebraic scaling consistent with finite-temperature calculations $\tilde{I}_{S}$. However, at very low temperatures, $\tilde{I}_{S,{\rm G}}$ starts to deviate from the scaling behavior. We attribute this deviation to strong finite-size effects (Friedel oscillations) inherent in the ground-state calculations, as demonstrated in Fig.~\ref{FigS2}(b,c). 

In Tab.~\ref{tab:Comp}, we summarize and compare the three approaches, $\tilde{I}_{S}$, $\tilde{I}_{S,G}$, and $\tilde{I}_2$, for evaluating the normalized spin current.

\begin{figure*}[ht]
\includegraphics[width=0.9\linewidth]{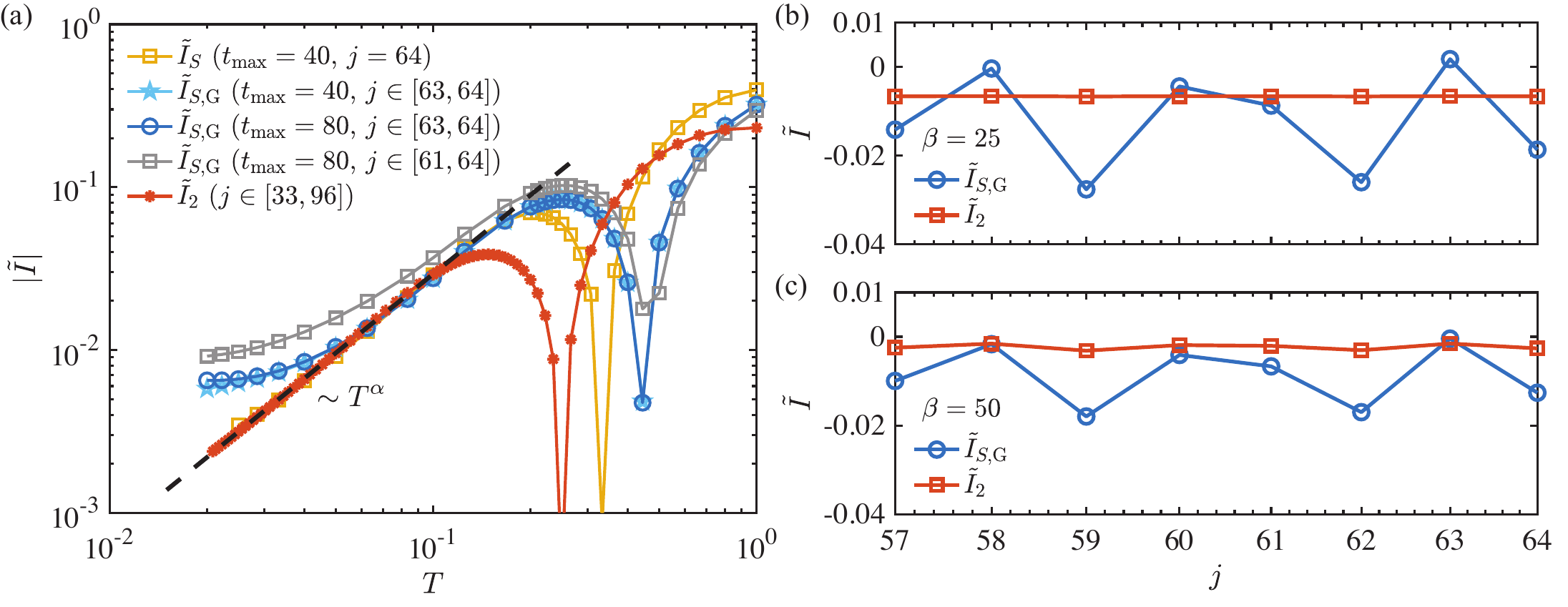}
\caption{ {(a) Comparison of the spin current results obtained via different methods on a spin chain of length $L=128$ and under magnetic field $B/J=1$. We retain bond dimension $D=500$ to ensure well converged results for all three quantities. The value of $\tilde{I}_S$ is computed at the central site $j=64$, while the two $\tilde{I}_{S, {\rm G}}$ curves are averaged over the adjacent sites $j\in[63, 64]$ and $j\in[61, 64]$, respectively. The results of $\tilde{I}_2$ is estimated on the bulk of chain, i.e., averaged over $j \in [33,96]$. The dashed line represents $T^\alpha$ with $\alpha \simeq 1.59(2)$. A real-time evolution duration of $t_{\text{max}} = 40$ to $80$ is sufficient to obtain converged dynamical results. (b,c) Site distribution of the spin currents $\tilde{I}_{S, {\rm G}}$ and $\tilde{I}_2$ at two different temperatures ($\beta=25$ and 50), revealing stronger finite-size effects in $\tilde{I}_{S, {\rm G}}$ while more uniform distribution for $\tilde{I}_2$.}}
\label{FigS2}
\end{figure*}
}

\section{Analytical calculations of spinon spin current in 1D Tomonaga-Luttinger liquid}
We consider the spin-$\frac{1}{2}$ Heisenberg spin chain with $J_{xy}=J_{z}=1$ as the energy unit. The normalized spin current is determined by the imaginary part of the local dynamical susceptibility. To compare with our numerical results  {in the bulk of the chain}, we consider the periodic boundary condition and the bulk contributions of the dynamical susceptibility to the spin currents.  {For a realistic experimental setup, edge contributions are responsible for a second sign reversal at a lower temperature, as shown in Fig.~\ref{Fig2}(b) of the main text and Fig.~\ref{FigE1} in the Appendix. Our analytical calculation below considers only the bulk contribution of the spin chain and therefore cannot reproduce the second sign reversal. Predictions of this lower-temperature sign reversal, which requires the inclusion of open-boundary conditions, can be found in Ref.~\cite{wang2025}.}

For $0 \leq B < 2$ the ground state is the Tomonaga-Luttinger liquid (TLL) phase. Using the bosonized representation of the spin Hamiltonian, we arrive at a low-energy effective Hamiltonian given as~\cite{takayoshi2010coefficients} 
\begin{equation}
\label{eq:Heff}
 \mathcal{H_{\text{eff}} }=\int dx\frac{v}{2}\{K^{-1}[\partial _{x}\phi(x)]^{2}+K[\partial _{x}\theta(x)]^{2}\},
\end{equation}
where $\phi(x)$ and $\theta(x)$ refer to the dual scalar fields, $K$ and $v$ refer to the TLL parameter and spinon velocity, respectively. The $\cos[\sqrt{16\pi}\phi(x)]$ term is irrelevant at finite magnetic fields~\cite{hikihara2004correlation}, thus is ignored in Eq.~(\ref{eq:Heff}).

The TLL parameter $K$ is related to the compactification radius $R$ via $K=1/(4\pi R^{2})$. However, $R$ and $K$ are only explicitly solvable when $B=0$ and $2$. To obtain their values for $0<B<2$, we follow the procedure in Ref.~\cite{cabra1998magnetization}; also see the references within Ref.~\cite{cabra1998magnetization}. First, a dressed energy function $\varepsilon_{d}(\eta)$ is introduced and solved using the integral equation of

\begin{equation}
\label{eq:integral_energy}
\varepsilon_{d}(\eta)=B-\frac{2}{\eta^{2}+1}-\frac{1}{2\pi}\int_{-\Lambda}^{\Lambda}\frac{4}{(\eta-\eta')^{2}+4}\varepsilon_{d}(\eta')d\eta'
\end{equation}
where the real positive parameter $\Lambda$ is determined by the condition of $\varepsilon_{d}(\Lambda)=0$. In the limit of $B=0$, $\Lambda=\infty $, and for $B\ll 1$ an approximate expression is also given in Ref.~\cite{bogoliubov1986critical}. After determining the value of $\Lambda$, a dressed charge function $\xi(\eta)$ is introduced as the solution of another integral equation given as

\begin{equation}
\label{eq:integral_charge}
\xi(\eta)=1-\frac{1}{2\pi}\int_{-\Lambda}^{\Lambda}\frac{4}{(\eta-\eta')^{2}+4}\xi(\eta')d\eta'
\end{equation}
where the compactification radius $R$ is determined by $R=1/(\sqrt{4\pi }\xi(\Lambda))$, or equivalently we can obtain $K=\xi(\Lambda)^{2}$.

Then, we turn to the dynamical spin susceptibility at finite temperatures. The large distance behavior of the dynamical spin susceptibility is carried out by combining the Bethe-Ansatz results and field theories. The spectral weight is most dominant when the momentum is near $\pi$ due to antiferromagnetic couplings. Following Ref.~\cite{Hirobe2017}, the expression of the dynamical spin susceptibility $\chi^{-+}(\pi + q,\omega)$ is given as 

\begin{align}
\label{eq:DSS}
\chi^{-+}(\pi +q,\omega)= \Theta(T,K) B(\frac{1}{8K}-i\frac{\omega-vq}{4\pi T},1-\frac{1}{4K})B(\frac{1}{8K}-i\frac{\omega+vq}{4\pi T},1-\frac{1}{4K})
\end{align}
where $v$ is the spinon velocity and $\Theta(T,K)$ is determined by 

\begin{align}
\label{eq:DSS_Theta}
\Theta(T,K)= -2A_{x}(K)\frac{(2-\frac{1}{K})\sin(\frac{\pi}{4K})}{\sin(\frac{\pi}{2K})}(\frac{\sin(\frac{\pi}{2K})}{2\pi T(2-\frac{1}{K})})^{2-\frac{1}{2K}}.
\end{align}

In Eq.~(\ref{eq:DSS_Theta}), the nonuniversal amplitude $A_{x}(K)$ is related to $B_{0}(K)$ in Eq. (S6) of Ref.~\cite{Hirobe2017} by the equation of $A_{x}(K)=B_{0}^{2}(K)/2$~\cite{hikihara2004correlation}; see more detailed expression of $A_{x}(K)$ in Ref.~\cite{bocquet2001finite}. With Eq.~(\ref{eq:DSS}) we can obtain the temperature dependence of $\chi^{-+}(\pi +q,\omega)$, which is valid for small $q$, low energies $\omega$, and low temperatures $T$.

However, Eq.~(\ref{eq:DSS}) assumes the linear spinon dispersion with spinon velocity $v$. In this approximation ${\rm Im}[\chi^{-+}(\pi +q,\omega)]$ is an odd function of $\omega $, leading to a zero normalized spin current for any magnetic field. For larger magnetic fields, the nonlinear spinon dispersion becomes important in the excitation spectrum, and leads to some corrections to Eq.~(\ref{eq:DSS}). To fully consider the nonlinearity of the dispersions, one needs to start from the nonlinear TLL theory~\cite{imambekov2012one}, which is beyond the scope of this paper. Here we follow the Supplementary Information of Ref.~\cite{Hirobe2017}. The linear terms $\pm vq$ in Eq.~(\ref{eq:DSS}) are replaced by the nonlinear dispersion $-\epsilon(\mp q)$, which is determined by the lower boundary of the spinon excitation continuum near $\pi $. The $\epsilon(q)$ under a finite magnetic field is given as~\cite{muller1981quantum}

\begin{align}
\label{eq:DSS_dispersion}
\epsilon(q)=2[\frac{\pi}{2}+\frac{B}{2}(1-\frac{\pi}{2})]\cos(\frac{q}{2})\sin(\frac{q}{2}+\pi M)-B
\end{align}
where $M=\frac{1}{\pi}\sin^{-1}(\frac{1}{1-\pi/2+\pi/B})$ is the approximate analytical expression for the magnetization associated with Eq.~(\ref{eq:DSS_dispersion}).

Finally, the normalized spin current is calculated by integrating the imaginary part of the $\chi^{-+}(\pi + q,\omega)$ over $\omega$ and $q$. In practice, a cutoff $\omega_{\rm max}$ is introduced in the integration for calculations at low temperatures. Because of the kernel function in the formula for the spin current, we find that the integrand becomes neglectable for $\omega > \omega_{\rm max}$. For example, at $B=1$, for $T<0.01$ it is sufficient to choose $\omega_{\rm max}=0.2$. The local dynamical spin susceptibility is obtained by integrating over $q$. In our calculations, a cutoff $q_{\rm max}$ is also used and determined by the corresponding $\omega_{\rm max}$ in the spectrum to make sure that the dynamical spin susceptibility given in Eq.~(\ref{eq:DSS}) remains valid within the ranges.

\begin{figure*}[htpb]
\includegraphics[width=1\linewidth]{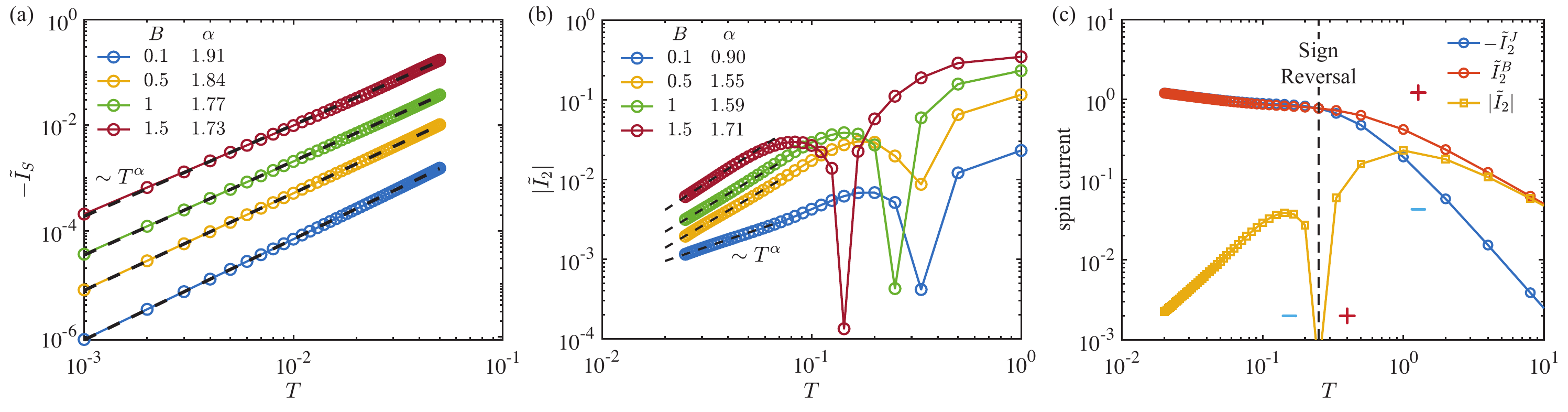}
\caption{(a) Analytical result of the spin current $\tilde{I}_S$ and (b) Numerical result of the spin current $\tilde{I}_2$ at selected magnetic fields within the TLL regime. Black dashed lines indicate $T^\alpha$ power-law fits. (c) Simulated spin current and its components of 1D Heisenberg chain with $\tilde{I}_2=\tilde{I}_2^J+\tilde{I}_2^B$, computed under a magnetic field $B=1$. The vertical black dashed line indicates the location of sign reversal, where the $(-\tilde{I}_2^J)$ and $\tilde{I}_2^B$ lines cross. A bond dimension of $D = 500$ is retained in the calculations.}
\label{FigS3}
\end{figure*}

We show the temperature dependence of the spin current $\tilde{I}_{S}$ at finite magnetic fields in Fig.~\ref{FigS3}(a). The $\tilde{I}_{S}$ results are negative and exhibit algebraic decay at low temperatures, consistent with our numerical results in the TLL phase. However, we notice that the exponents $\alpha$ do not agree with the numerical calculations [see Fig.~\ref{FigS3}(b)], especially in the low magnetic field regime where higher orders of the nonlinear spinon dispersions cannot be ignored. In Fig.~\ref{FigS3}(c), we show the decomposition of spin current $\tilde{I}_2 = \tilde{I}_2^J + \tilde{I}_2^B$, demonstrating sign reversal of net spinon spin current due to competition between interaction ($\tilde{I}_2^J$) and Zeeman-term ($\tilde{I}_2^B$) contributions.

\section{Sign Correspondence Between Spin Current and Magnetization Derivative}
\begin{figure}[htpb]
\includegraphics[width=1\linewidth]{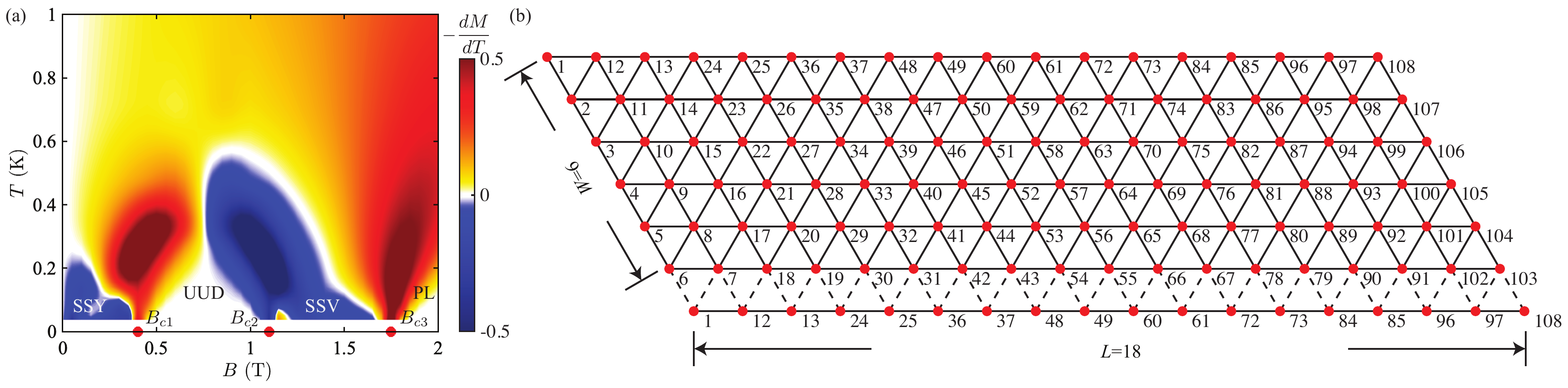}
\caption{(a) Calculated $-\frac{dM}{dT}$ results for the realistic easy-axis TLAF model with $D=3000$. (b) The Y-type cylinder used in the calculation with width $W=6$ and length $L=18$ (YC$6\times18$). The black dashed lines indicate the periodic boundary condition along $y$ axis. Magnetization $M$ is computed in the bulk region (19-90) to minimize finite-size effects.
}
\label{FigS4}
\end{figure}

 {Our results reveal a correspondence in sign between the spin current and the temperature derivative of the magnetization, $dM/dT$.} Assuming $T_s < T_m$ ($\delta T < 0$), we find the spin current direction is positive (outflow) when $-\frac{dM}{dT} > 0$ ($\delta M < 0$) and negative (inflow) when $-\frac{dM}{dT} < 0$ ($\delta M > 0$). Figure~\ref{FigS4}(a) confirms such correspondence numerically in the spin supersolid phase, showing alignment between the sign of $-\frac{dM}{dT}$ and the normalized current $\tilde{I}_2$, establishing the derivative as a  indicator of spin-current direction. The calculations are conducted on Y-type cylinder (of size YC$6\times18$) as illustrated in Fig.~\ref{FigS4}(b).

\section{Linear spin-wave theory for spin Seebeck effect}
Now we apply the linear spin-wave theory (LSWT) to the easy-axis triangular-lattice antiferromagnetic model  {under a magnetic field}
\begin{equation}
H = J \sum_{\langle i,j \rangle} (S^x_i S^x_j + S^y_i S^y_j + \Delta S^z_i S^z_j) - B \sum_i S_i^z,
\end{equation}
where  {$J\equiv 1$} is the energy scale and $\Delta$ is the anisotropic parameter. Under the linear spin-wave approximation, the ground states of the model with $\Delta > 1$ is a Y-shaped supersolid state  {(SSY)}, a up-up-down solid state (UUD), a V-shaped supersolid state  {(SSV)} and a polarized state, separating by three quantum critical point with 
\begin{equation}
\begin{split}
B_{c1} &= 3 S,\\
B_{c2} &= 3 S (\Delta - \frac{1}{2} + \sqrt{\Delta^2 + \Delta - \frac{7}{4}}),\\
B_{c3} &= 3 S (1 + 2 \Delta).
\end{split}
\end{equation} 

 {By assuming three-sublattice spin order within the $x$-$z$ plane}, we introduce three kinds of Holstein-Primakoff bosons $a_{1,2,3}$ on sublattices $A_{1,2,3}$ to parametrize the spin operators,  {i.e.,}
\begin{equation}
\begin{split}
	&S^z_n =\cos\theta_n ~ (S-a_n^\dagger a_n)- \sin \theta_n ~\frac{\sqrt{2S}}{2}(a_n+a_n^\dagger), \\
	&S^x_n=\sin\theta_n ~ (S-a_n^\dagger a_n)+\cos \theta_n ~\frac{\sqrt{2S}}{2}(a_n+a_n^\dagger),\\
	&S^y_n=\frac{\sqrt{2S}}{2i}(a_n-a_n^\dagger),
\end{split}
\end{equation}
where $\theta_n$ can be obtained by minimizing the classical energy
\begin{equation}
E =  \frac{S^2}{2} \sum_{n \neq n'} \sin \theta_n \sin\theta_{n'} + 
\frac{\Delta S^2}{2} \sum_{n \neq n'} \cos \theta_n \cos\theta_{n'}
- \frac{BS }{3} \sum_n \cos \theta_n.
\end{equation}

Now we consider the interactions between $n$ and $n'$ sites (only two operator terms):
\begin{equation}
\begin{split}
S^xS^x:~ & -S \sin\theta_n \sin\theta_{n'} (a_n^\dagger a_n + a_{n'}^\dagger a_{n'})
+\frac{S}{2} \cos\theta_n \cos\theta_{n'} (a_n a_{n'} + a_n a_{n'}^\dagger + a_n^\dagger a_{n'}
+a_n^\dagger a_{n'}^\dagger ),\\
S^yS^y:~ & -\frac{S}{2} (a_n a_{n'} - a_n a_{n'}^\dagger - a_n^\dagger a_{n'}
+a_n^\dagger a_{n'}^\dagger ),\\
\Delta S^zS^z:~ & -S \Delta \cos\theta_n \cos\theta_{n'} (a_n^\dagger a_n + a_{n'}^\dagger a_{n'})
+\frac{S\Delta}{2} \sin\theta_n \sin\theta_{n'} (a_n a_{n'} + a_n a_{n'}^\dagger + a_n^\dagger a_{n'}
+a_n^\dagger a_{n'}^\dagger ).\\
\end{split}
\end{equation} 
By introducing the Fourier transformation $a_{n,i} = \frac{1}{\sqrt{N}} \sum_k e^{ik r_{n,i}} a_{n,k}$, we arrive at the quadratic Hamiltonian in momentum space $$H_k = \sum_k \alpha_k^\dagger H_0(k) \alpha_k,$$ with $\alpha_k^\dagger  = (a_{1,k}^\dagger~a_{2,k}^\dagger ~a_{3,k}^\dagger~a_{1,-k}~a_{2,-k}~a_{3,-k})$.
Now we perform Bogoliubov transformation to diagonalize $H_k$, i.e., find a matrix $Q$ such that $(Q^{-1})^\dagger H_0(k)Q^{-1}=\hat{\lambda}\equiv\diag[\lambda_1,\lambda_2,\lambda_3,-\lambda_1,-\lambda_2,-\lambda_3]$. To maintain the bosonic commutation relation for $\beta_k^\dagger = (b_{1,k}^\dagger~b_{2,k}^\dagger ~b_{3,k}^\dagger~b_{1,-k}~b_{2,-k}~b_{3,-k})$ defined via $\beta_k=Q\alpha_k$, the transformation $Q$ must satisfy $QLQ^\dagger = Q^\dagger L Q=L$, $Q^\dagger L=L Q^{-1}$, with $L=\diag[1,1,1,-1,-1,-1]$. \\

In practice, we follow the four steps below to determine $Q$:
\begin{itemize}
\item (i) Find $K$ such that $H_0(k)=K^\dagger K$;
\item (ii) Diagonalize $KLK^\dagger$ with unitary matrix $U$ such that $U^\dagger (KLK^\dagger)U=\hat{\lambda}$;
\item (iii) Obtain the eigenvalue $\lambda=L\hat{\lambda}$;
\item (iv) Obtain the transfer matrix $Q=(\sqrt{\lambda})^{-1}U^\dagger K$.
\end{itemize}

\begin{figure}[htpb]
\includegraphics[width=1\linewidth]{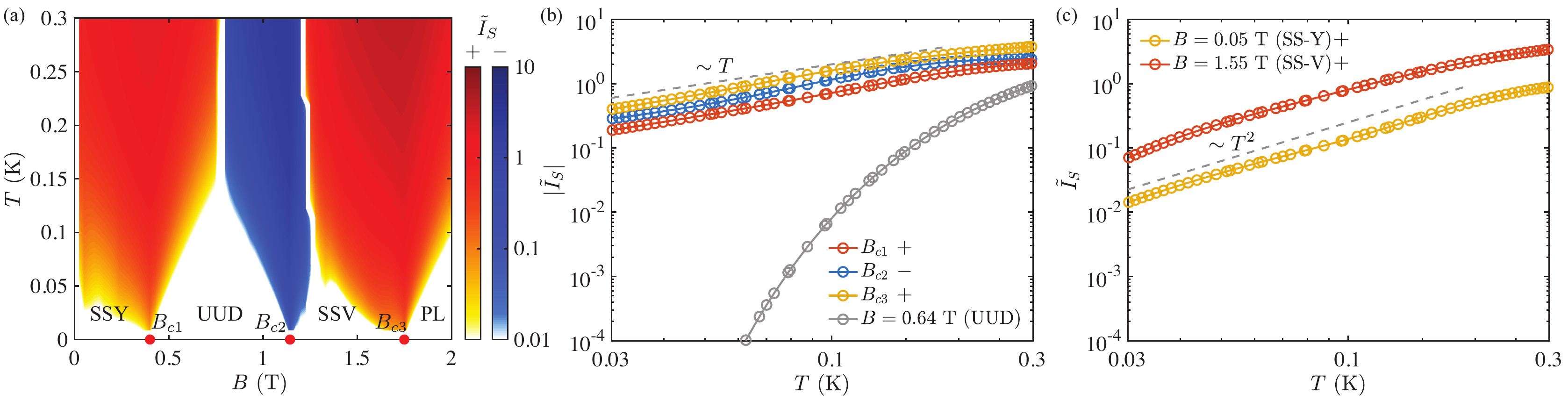}
\caption{ {(a) The LSWT results for the spin current $\tilde{I}_S$ (with the realistic model parameter of NBCP), which agree with tensor-network predictions in both high-temperature regime and near QCPs, but fail to capture the persistent supercurrent in the supersolid phase. The SSY, UUD, SSV, and PL label the same quantum spin states as in Fig.~3(a) in the main text, separated by three QCPs at $B_{c1,2,3}$. The red and blue color bars represent the positive and negative spin currents, respectively.}
(b) The LSWT results of spin current $\tilde{I}_S$ near three critical fields and in the UUD phase. 
(c) Results in the SSY and SSV phases,  {where no sign reversal is observed.} 
The ``$+$'' and ``$-$'' signs  {in the legends of all panels} represent the positive and negative spin current, respectively.
}
\label{FigS5}
\end{figure}

Within the single-magnon Hilbert space, we can obtain the local dynamical susceptibility
\begin{equation}
\begin{split}
\frac{1}{2}{\rm Im}[\chi^{-+}_{\rm loc} (\omega) + \chi^{-+}_{\rm loc} (-\omega)] &=
\frac{\pi}{2 \mathcal{Z}}\sum_{m,n}(|| \langle m| S_j^- |n \rangle ||^2 - || \langle m| S_j^+ |n \rangle ||^2)
e^{-\beta E_n} (1-e^{-\beta \omega}) \delta(\omega + E_n - E_m)\\
&\simeq \frac{\pi}{2N}\sum_{m,k,i}(|| \langle 0| b_{i,k}S_k^- |0 \rangle ||^2 - || \langle 0| b_{i,k}S_k^+ |0 \rangle ||^2) (1-e^{-2\beta \lambda_i}) \delta(\omega -2\lambda_i),
\label{EqS:SqChi}
\end{split}
\end{equation}
with $S_k^+=\sum_{i=1}^3 \frac{1}{2}\cos \theta_n (a_{i,k}+a_{i,k}^\dagger)+\frac{1}{2}(a_{i,k}-a_{i,k}^\dagger)$, 
$a_{j,k}=\sum_{l=1}^3P_{j,l} b_{l,k}+\sum_{l=1}^3P_{j,l+3} b_{l,-k}^\dagger$, and $P=Q^{-1}$.
Thus we have $\langle0|b_{i,k}a_{j,k}|0\rangle=P_{j,i+3}$, $\langle0|b_{i,k}a_{j,k}^\dagger|0\rangle=P_{j,i}^*$, and 
\begin{equation}
\begin{split}
    || \langle 0| b_{i,k}S_k^+ |0 \rangle ||^2 
    &= || \langle 0| b_{i,k}
    \sum_{l=1}^3 \frac{1}{2}(\cos \theta_l+1)a_{l,k} + \frac{1}{2}(\cos \theta_l-1)a_{l,k}^\dagger |0 \rangle ||^2\\
    &= || 
    \sum_{l=1}^3 \frac{1}{2}(\cos \theta_l+1)P_{l,j+3} +  \frac{1}{2}(\cos \theta_l-1) P_{l,j}^*||^2; \\
    || \langle 0| b_{i,k}S_k^- |0 \rangle ||^2 
    &= || \langle 0| b_{i,k}
    \sum_{l=1}^3 \frac{1}{2}(\cos \theta_l-1)a_{l,k} + \frac{1}{2}(\cos \theta_l+1)a_{l,k}^\dagger |0 \rangle ||^2\\
    &= || 
    \sum_{l=1}^3 \frac{1}{2}(\cos \theta_l-1)P_{l,j+3} +  \frac{1}{2}(\cos \theta_l+1) P_{l,j}^*||^2.
    \label{EqS:Ele}
\end{split}
\end{equation}
 
By substituting Eq.~(\ref{EqS:Ele}) into Eq.~(\ref{EqS:SqChi}), we compute $\tilde{I}_S$ within LSWT,  {with the results shown in Fig.~\ref{FigS5}}. 
While the linear-$T$ behavior at QCPs agrees with tensor-network results in the main text (Fig.~3(c)), LSWT exhibits significant limitations in the supersolid phase. Specifically, it predicts a $T^2$ temperature dependence [Fig.~\ref{FigS5}(c)] rather than the persistent currents observed in tensor-network numerical simulations, and produces exclusively positive currents in both SSY and SSV phases [Fig.~\ref{FigS5}(a)] --- in stark contrast to the behavior shown in Fig.~4(a). These discrepancies clearly indicate that investigating spin currents in the supersolid phase necessitates theoretical approaches beyond conventional LSWT,  {like the tensor-network approach for $\tilde{I}_2$ developed here, to accurately capture these quantum spin transport behaviors.}

\end{document}